\documentclass[universe,review,accept,pdftex,moreauthors]{Definitions/mdpi} 
\usepackage{aasmacros}

\usepackage{color}

\firstpage{1} 
\pubvolume{1}
\issuenum{1}
\articlenumber{0}
\pubyear{2024}
\copyrightyear{2024}
\datereceived{31 January 2024} 
\daterevised{13 March 2024} 
\dateaccepted{15 March 2024} 
\datepublished{ } 
\hreflink{https://doi.org/10.3390/universe10040163} 

\usepackage{textcomp} 

\Title{A Very-High-Energy Gamma-Ray View of the Transient Sky  }

\TitleCitation{A Very-High-Energy Gamma-Ray View of the Transient Sky}

\Author{Alessandro Carosi~$^{1,}$*$^{,\dagger}$\orcidA{} and Alicia L\'opez-Oramas~$^{2,}$*$^{,\dagger}$\orcidB{}}

\AuthorNames{Alessandro Carosi and Alicia L\'opez-Oramas}

\AuthorCitation{Carosi, A.;  L\'opez-Oramas, A.}

\address{%
$^{1}$ \quad INAF---Osservatorio Astronomico di Roma, Via Frascati 33, I-00040
Monte Porzio Catone, Italy\\
$^{2}$ \quad Instituto de Astrof\'isica de Canarias and
 Departamento de Astrof\'isica, Universidad de La Laguna, \mbox{38200 La Laguna}, Tenerife, Spain}

\corres{Correspondence: alessandro.carosi@inaf.it (A.C.); alicia.lopez@iac.es (A.L.-O.)}

\firstnote{These authors contributed equally to this work.}

\abstract{The development of the latest generation of Imaging Atmospheric Cherenkov Telescopes (IACTs) over recent decades has led to the discovery of new extreme astrophysical phenomena in the very-high-energy (VHE, E~>~100~GeV) gamma-ray regime. Time-domain and multi-messenger astronomy are inevitably connected to the physics of transient VHE emitters, which show unexpected (and mostly unpredictable) flaring or exploding episodes at different timescales. These transients often share the physical processes responsible for the production of the gamma-ray emission, through cosmic-ray acceleration, magnetic reconnection, jet production and/or outflows, and shocks interactions. In this review, we present an up-to-date overview of the VHE transients field, spanning from  novae to supernovae, neutrino counterparts or fast radio bursts, among others, and we outline the expectations for future facilities.}


\begin{document}


\section{Introduction}

The very-high-energy (VHE, E $>$  100~GeV) gamma-ray regime is of the utmost importance in studying extreme astrophysical processes. Transient phenomena, located at the crossroads of time-domain and multi-messenger astronomy, have revealed a plethora of new emitters at VHE. During the last twenty years, IACT experiments have proven to be suitable instruments to perform fast follow-up of transient events, with 3--4 times improved sensitivities at short time scales compared to space-based instruments~\citep{Fioretti2019ICRC...36..673F}. At the same time, some IACTs are optimized for a swift reaction and repositioning,~see e.g.,~\citep{2016APh....72...61A}, allowing for the study of short-lived signals during their initial phase. Understanding the recent advancements and open issues in transient and multi-messenger astrophysics at VHE is the key for the science to be developed with current IACTs and with future instrumentation, such as the Cherenkov Telescope Array (CTA) Observatory \citep{CTA2019scta.book.....C}. 

In this contribution, we review the phenomenology of transient events of both Galactic and extragalactic origin, which are (mostly) related to stellar-size compact objects and that are shock-powered and/or accretion-powered. Shocks power several transient phenomena, such as novae, supernovae, binary neutron star mergers, or tidal disruption events. The shocks and ejecta in (at least some of) these systems can show morphological resemblances and have similar characteristics, although at different scales and displaying, e.g., a broad range of various velocities and densities. Other sources such as magnetars, even if rotationally powered, can also generate blast waves and shocks. 

We review the state of the art in the detection and search for emission of transient events in the~GeV-TeV regime for different types of astrophysical sources in \mbox{Sections~\ref{sec_novae}--\ref{sec_other}}, namely novae (Section~\ref{sec_novae}), microquasars, and flaring gamma-ray binaries (Section~\ref{sec_mic}), supernovae (Section~\ref{sec_sne}), pulsar-wind nebulae (Section~\ref{sec_pwne}), fast radio bursts and magnetars (Section~\ref{sec:frbs}), and gravitational waves (Section~\ref{sec_gws}). We then briefly mention other transient sources in Section~\ref{sec_other}, such as gamma-ray bursts (GRBs) and tidal disruption events (TDEs), which are more extensively covered in a dedicated review of this Special Issue~\citep{2024VL}. We finally discuss the latest advances and future perspectives in Section~\ref{sec_discussion}.

\section{Novae}
\label{sec_novae}

Novae outbursts are thermonuclear explosions that take place on the surface of a white dwarf (WD) accreting material from a companion star. The transferred material is mostly hydrogen; however, helium accretion can also occur in some systems (see~\citep{Chomiuk2021ARA&A..59..391C}). The material accumulates on a layer on the surface of the WD, where hydrogen is burning in degenerate conditions, increasing the temperature and density. Once a critical mass is reached, the system undergoes an unstable burning, provoking a thermonuclear runaway. The ejecta expand at velocities reaching thousands of km s$^{-1}$ and can interact with the surrounding material, if any. Depending on the type of companion star, novae can be classified into \textit{classical} and \textit{symbiotic} systems. Classical novae are cataclysmic variables in which the companion is a main sequence (or slightly evolved) star. The mass-transfer onto the WD happens via Roche-lobe overflow. These systems are characterized for having short orbital periods lasting from hours to days \citep{Hernanz2012BaltA..21...62H}. Symbiotic systems are composed of a red giant (RG) companion and the have longer orbital periods and show larger component separations \citep{Mikolajewska2008ASPC..401...42M}. The binary is embedded in the RG wind and the WD accretes directly from this wind. 

Novae explosions do not disrupt the binary system and hence the cycle of accretion can start again. After enough material is accumulated, another thermonuclear runaway can happen again. The recurrence timescale of these outburst is defined as $\tau_{rec} = M_{acc} / \dot{M}$, $M_{acc}$ being the critical mass to initiate the nuclear burning and $\dot{M}$ the accretion rate. High recurrence times are then associated to more massive WDs (with mass close to the Chandrasekhar limit, $M_{Ch}$ $\approx 1.4 M_{\odot}$) accreting at high rates. The typical recurrence times for classical novae are $10^{4}$--$10^{5}$ years. However, some systems known as \mbox{\textit{recurrent novae}} have displayed more than one eruption in a human lifetime ($\tau_{rec} \le 100$ years). For this to happen, the WD should be close to the Chandrasekhar limit (with at least $M > 1.2 M_{\odot}$) and the system shall have high mass-accretion rates ($\approx$10$^{-7} M_{\odot}$ y$^{-1}$) (see~\citep{Schaefer2010ApJS..187..275S}). There are 10 recurrent systems known in the Galaxy up to date  which can be further classified into two groups~\citep{Schaefer2010ApJS..187..275S} depending on the mechanism which leads to the short recurrence: long-period systems (eight in total, period  $>$ one-third of a day) hosting a giant companion (also known as \textit{symbiotic recurrent novae}) in which the accretion is driven by the RG wind and the evolution of the companion. These symbiotic recurrent systems can indeed be the progenitors of type Ia supernovae; and short-period ones (two systems) in which the accretion is driven by the heating of the WD. 

The discovery rate of novae is about 5--15 events per year~\citep{Chomiuk2021ARA&A..59..391C}, although the estimated number of eruptions in the Galaxy is much larger: 20--70 per year. The lower detection rate could be due to dust obscuration, since many novae happen in the Galactic plane, or simply due to a scarce monitoring;~see, e.g.,~\citep{Schaefer2010ApJS..187..275S}, and references therein for a more detailed discussion on novae rates.

The first evidence of non-thermal emission due to particle acceleration up to TeV energies in the blast wave of (recurrent) novae was suggested by~\cite{Tatischeff2007ApJ...663L.101T}. The discovery of symbiotic novae as high energy (HE; E  $>$  100 MeV) emitters was performed by \textit{Fermi}-LAT in 2010~\citep{Fermi2010Sci...329..817A} with the detection of V407 Cyg. The HE emission lasted for about two weeks. Only four years later, \textit{Fermi}-LAT also established classical novae as HE sources~\cite{Fermi2014Sci...345..554A} with the discovery of three systems (V959 Mon, V1324 Sco, and V339 Del). The spectral energy distribution (SED) of these four LAT-detected novae is rather soft, mostly described with power laws with exponential cutoff and with energies up to a maximum of $\sim$10~GeV. Both hadronic and leptonic scenarios can fit the observed emission and could not be ruled out. 

Since then, the satellite has been detecting an average of $\sim$1 nova per year.\endnote{See \url{https://asd.gsfc.nasa.gov/Koji.Mukai/novae/latnovae.html} (accessed on 27 March 2024) for the list with LAT-detected novae and sub-significance hints.} Most of the \textit{Fermi}-LAT novae are located in the Galactic disk, although some have been discovered in the Galactic bulge, implying detection up to distances of $\sim$8 kpc. By studying different classical novae,~ref.\cite{Cheung2016ApJ...826..142C} suggests an inverse relationship between the HE emission duration and the total emitted energies. This could possibly indicate that the presence of more compact and high-density ejecta produces a higher particle acceleration, which leads to stronger emissions and shorter duration.

The detection of HE emission from novae clearly demonstrated that non-thermal mechanisms operate in these cataclysmic binaries. The evident question to pose was whether novae could accelerate particles to sufficiently high energies to produce VHE gamma rays. These particles (leptons and protons) are accelerated at the nova shock and could eventually produce emission at higher energies. In the case of protons, they could reach high energies and emit TeV gamma rays~\citep{Sitarek2012PhRvD..86f3011S}. Since the discovery of HE gamma rays in novae, searches for a VHE component were performed by IACTs for over a decade, without achieving any significant detection. VERITAS observed the 2010 outburst of the symbiotic nova V407 Cyg on days 9 to 16 after the eruption, leading to no signal~\citep{Aliu2012ApJ...754...77A}. MAGIC observed the classical nova V339 Del on the night of the optical peak (although under poor-quality weather conditions) and a few days after the \textit{Fermi}-LAT emission, setting upper limits (ULs) to the VHE emission~\citep{Ahnen2015A&A...582A..67A}. In the same work, MAGIC reported no signal from the symbiotic nova YY Her (taken a week after the optical maximum) and the dwarf nova ASASSN-13ax, a system in which the outburst are due to accretion disk instabilities (instead of a thermonuclear runaway). 

The first nova for which VHE gamma-ray emission was discovered is RS Oph, a recurrent symbiotic system composed of a massive $M_{WD} \approx 1.2\div$1.4 M$_{\odot}$) carbon--oxygen WD~\cite{Mikolajewska2017ApJ...847...99M} accreting from a M0-2 III  RG star~\cite{Anupama1999A&A...344..177A}.  It shows an orbital period of $(453.6 \pm 0.4)$ days~\cite{2009A&A...497..815B} and displays major outbursts with a recurrence time of 14.7 years~\cite{Schaefer2010ApJS..187..275S}. The fact that the mass of the WD is so close to the Chandrasekhar limit suggest that RS Oph is a possible type Ia SN (see Section~\ref{sec_sne}) progenitor candidate~\citep{Hachisu2000ApJ...536L..93H,Hernanz2008NewAR..52..386H,Patat2011A&A...530A..63P}.

\begin{figure}[H]
    \centering
    \includegraphics[width=10cm]{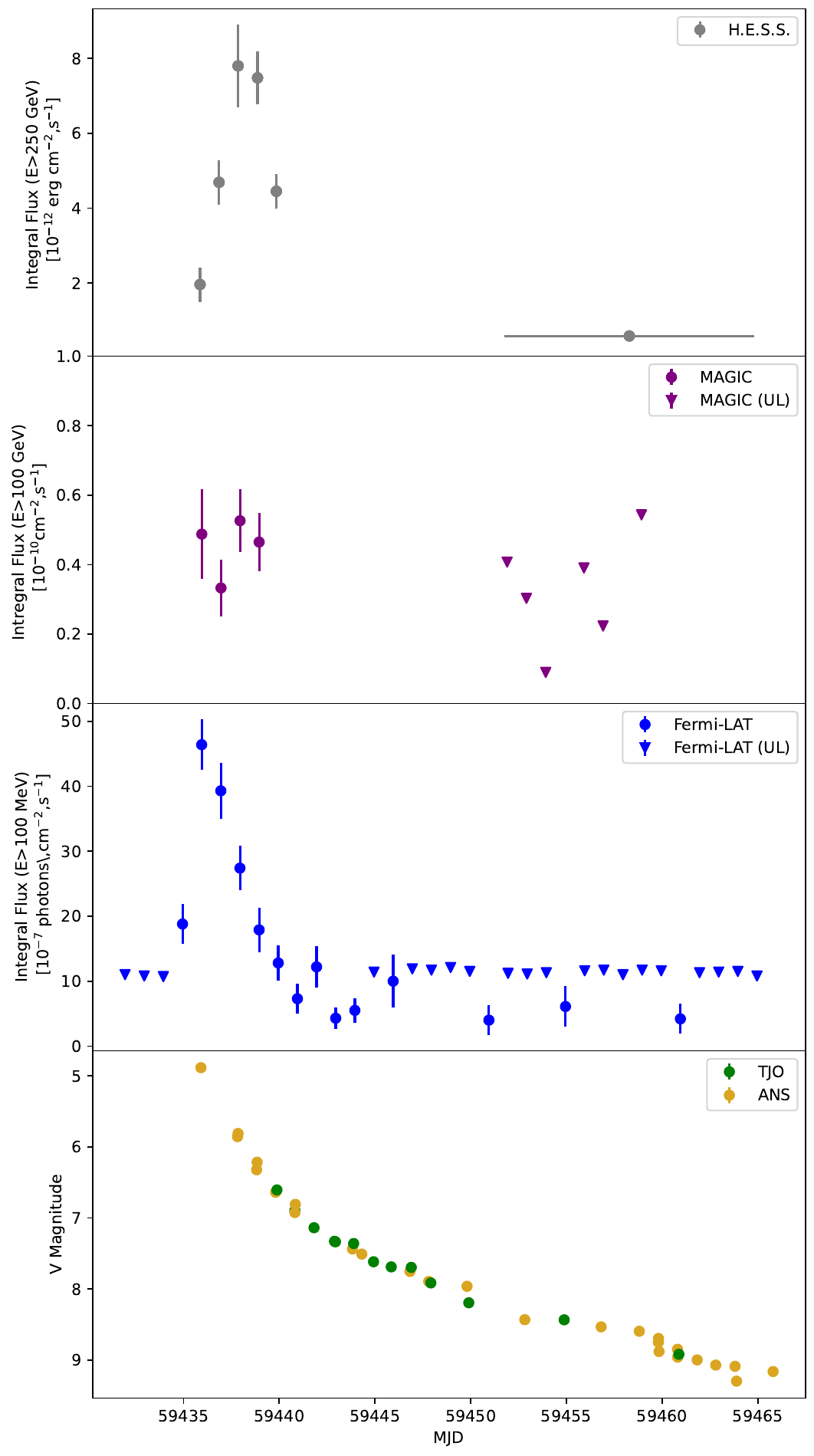} 
    
    \caption{Multi-wavelength lightcurve of the 2021 eruption of RS Oph as seen by   H.E.S.S (top panel), MAGIC (second panel), \textit{Fermi}-LAT, (third panel), and optical V magnitude (bottom) observations by ANS (golden points) and TJO (green points). ULs for the late MAGIC emission and \textit{Fermi}-LAT are indicated as inverted triangles. Optical, \textit{Fermi}-LAT, and MAGIC data points taken from the supplementary material in \cite{Acciari2022NatAs...6..689A}, H.E.S.S. fluxes from auxiliary material from \cite{hess2022Sci...376...77H}.   
    \label{fig_rsoph_mwl}}
\end{figure}

The VHE observations were triggered on 9 August, after optical~\cite{KGeary2021} and HE~\cite{2021ATel14834....1C} alerts. A clear gamma-ray signal at VHE was then detected by H.E.S.S~\citep{hess2022Sci...376...77H}, MAGIC~\citep{Acciari2022NatAs...6..689A}, and confirmed by the LST-1 telescope~\citep{Abe:2023QP} during the 2021 outburst that started on 8 August 2021 (MJD 59435). The multi-wavelength lightcurve of the RS Oph emission is shown in Figure~\ref{fig_rsoph_mwl}). The VHE gamma-ray signal is significantly detected up to five days after the nova eruption. Observations after the full Moon break revealed no significant signal, with a maximum of 3.3 $\sigma$ hint integrated over 14.6 h between 25 August and 7 September~\citep{hess2022Sci...376...77H}. 

The lightcurve reported by IACTs varies in shape depending on the energy range (see Figure~\ref{fig_rsoph_mwl}). MAGIC observed RS Oph between 9 August and 1 September, for a total of 21.4~h. The signal detected in the VHE regime measured during the first 4 days corresponds to the optical and HE maxima. However, the 4-day binned emission  $>$100~GeV is best fit to a constant flux~\citep{Acciari2022NatAs...6..689A}, which suggest a migration of the gamma-ray emission toward higher energies, implying an increase in the energies of the parent particle population. A constant flux compatible with that measured by MAGIC has been reported by the LST-1 during the first nights~\citep{Yukiho2023arXiv231009683K}. On the other hand, H.E.S.S. observed this between 9 August and 7 September. The signal was detected by H.E.S.S. at  $>$250~GeV peaks a day after the HE maximum, with a temporal decay of t$^{-(1.43\pm0.48)}$ compatible with what observed at HE, explaining the similarities in the lightcurve due to a common origin of the emission, in which the particles are accelerated at the external shock~\citep{hess2022Sci...376...77H}.

\begin{figure}[H]
\centering
\includegraphics[width=13cm]{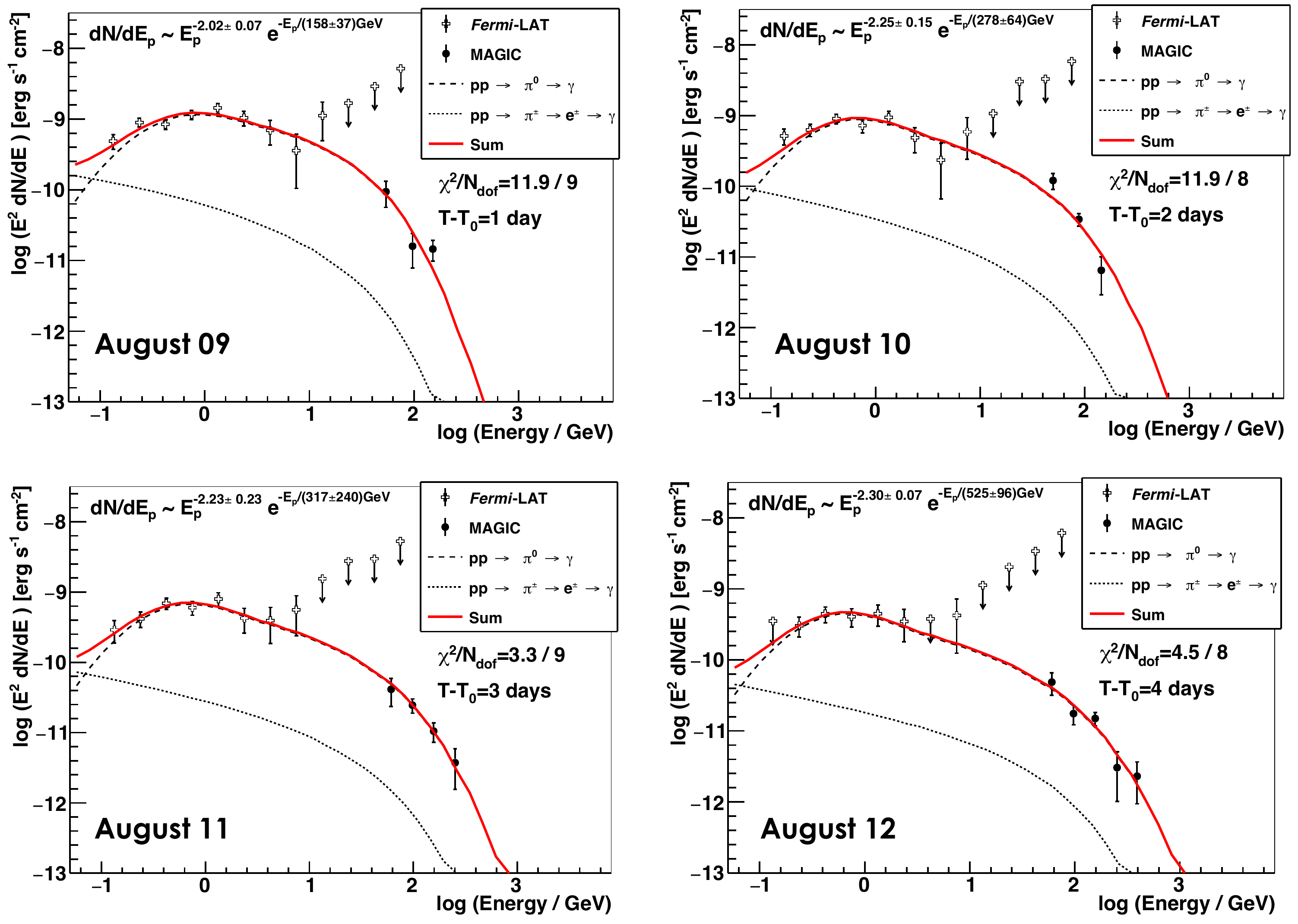}
\caption{Daily SEDs of RS Oph as seen by \textit{Fermi}-LAT and MAGIC adjusted to an hadronic scenario from the first night on 9 August 2021 to the fourth night, 12 August. A hint of spectral hardening is observed with increasing time, with increased cut-off energies. Reprinted with permission from \cite{Acciari2022NatAs...6..689A}.\label{fig2_rsoph}}
\end{figure}

The VHE component of the SED measured by MAGIC and H.E.S.S. expands from 60~GeV up to 1 TeV \cite{Acciari2022NatAs...6..689A,hess2022Sci...376...77H}. Both collaborations performed a joint analysis of the HE and VHE data and suggest that the combined \textit{Fermi}-LAT + MAGIC and \textit{Fermi}-LAT + H.E.S.S. spectra can be described as a single component spanning from 50 MeV to VHE. The emission would be due to a shock created by the ejecta which expand into the surrounding medium and the wind of the RG companion, creating a single shock where particles are accelerated. Both experiments suggest that the gamma-ray emission is best fit by a hadronic scenario, in which protons are accelerated in the shock wave formed by the interaction of the novae ejecta with  with the interstellar medium with some contribution of the RG wind. The daily SEDs (Figure~\ref{fig2_rsoph} for the MAGIC sample) are also best adjusted to a hadronic case, with evidence of increase in the energy cutoff, implying and acceleration of protons and the absence of strong cooling processes. The leptonic scenario does not properly fit the obtained spectra. A lepto-hadronic scenario is also tested by \citep{Acciari2022NatAs...6..689A}, providing a poor fit. In the case of protons, the injected particle spectrum also follows a canonical distribution ($\Gamma=-2$), while the leptonic and lepto-hadronic cases assume more complicated injection models with some strong ad hoc spectral breaks which cannot be fully explained by cooling, and still leading to a poorer fit of the SED. All this together favored the hadronic scenario as mechanism for the VHE gamma-ray production and the settlement of novae as proton accelerators. 

Ref.~\citep{DeSarkar2023ApJ...951...62D} elaborated a model to explain both the gamma-ray and radio emissions assuming a single shock, multi-population (lepto-hadronic) scenario for the first four nights of the outburst. The authors suggest a possible different origin for the HE and VHE components, the HE one being of leptonic origin and the VHE hadronic-dominated. This scenario would also explain the temporal dependence of the measured emissions. 

Novae ejecta are not spherical and show some asymmetries whose shape depends mainly on the densities of the surroundings. It has been proposed that, at least in classical novae, two components may be at work: a slow ($\sim$hundreds of km/s) dense flow (from binary motion) that moves in the equatorial plane and a faster ($\sim$few thousand of km/s) less dense isotropic outflow (wind from the WD) that propagates in the polar \mbox{direction~\citep{Chomiuk2014Natur.514..339C,Chomiuk2021ARA&A..59..391C}}. This creates a forward shock that is driven into the slow outflow, while a reverse shock will interact with the faster component. It is still unknown whether this scenario is universal to all types of novae or not.

The 2021 eruption of RS Oph could be a first step to answer this enigma. The 2006 outburst already showed some asymmetric structures in the ejecta with reported extended emission in the east--west direction \citep{Chesneau2007A&A...464..119C,Bode2007ApJ...665L..63B,Ribeiro2009ApJ...703.1955R}. A similar asymmetry has been reported during the latest outburst, displaying a bipolar structure with a predominant orientation in the same direction \citep{Munari2022A&A...666L...6M,Nikolov2023arXiv230911288N,Montez2022ApJ...926..100M}. Both in the 2006 and 2021 eruptions, a slow moving equatorial ring and a faster bipolar ejecta expanding in the polar (east--west) direction have been reported~\mbox{\citep{Chesneau2007A&A...464..119C,Nikolov2023arXiv230911288N}}. Ref.~\citep{Nikolov2023arXiv230911288N} claims the formation of a ring-like structure in the orbital plane due to the interaction between the ejecta and an enhanced ambient medium in the equatorial plane and argue that similar torus-like structures are observed in classical novae, although with a different origin (due to the orbital motion on the ejecta, see~\citep{Chomiuk2014Natur.514..339C}). 

The recent detection of the RS Oph during the 2021 outburst in the~GeV-TeV domain together with the multi-wavelength data have led some authors to suggest the presence of multiple shocks (polar and equatorial) in this system. Recently, ref.~\citep{Diesing2023ApJ...947...70D} argued that the presence of multiple ejecta components can explain both the gamma-ray SEDs and the shapes of the lightcurves as seen by \textit{Fermi}-LAT and H.E.S.S., which are consistent with the combination of two shocks: a low-velocity shock which expands in a dense medium and another faster one which expands in a less-dense environment. This scenario suggests that RS Oph (and probably RS Oph-like symbiotic systems) show similar properties to classical~novae. 

The protons that are accelerated in (symbiotic recurrent) novae ejecta can eventually escape onto the interstellar medium and contribute to the Galactic cosmic ray sea. Novae are less energetic that other events such as supernovae (SNe) ($\sim$$10^{43}$ erg vs. $\sim$$10^{50}$ erg) but they occur at a higher rate (5--15 detected novae in the Galaxy per year vs. 1--2 core-collapse SNe); hence, their contribution to the cosmic ray budget could be noticeable. However, considering the energetics of RS Oph as detected at VHE and its recurrence time of $\sim$15~years, the overall contribution to the cosmic ray spectrum is negligible, only 0.1\% of that of SNe \citep{Acciari2022NatAs...6..689A}. Nevertheless, novae do create bubbles of enhanced cosmic ray density in their close environment. In the case of RS Oph-like systems, these bubbles can extend up to $\sim$10~pc.

\section{Flaring Binary Systems and Microquasars}
\label{sec_mic}
Different types of gamma-loud binaries have been found to emit both in the HE and VHE regime, such as gamma-ray binaries and microquasars~\citep{Paredes2019RLSFN..30S.107P}. They represent a good opportunity to study particle acceleration in shocks and jets at relatively short timescales. The so-called gamma-ray binaries are systems that display the peak of their non-thermal emission above 1 MeV and are composed by massive stars (O or Be type) being orbited by a compact object (either NS or BH). Out of eight known gamma-ray binaries, three of them host a pulsar: PSR~B1259–63, PSR~2032+4127, and LS I +61 303. Even if the powering engine remains unknown for the others, it could be that all gamma-ray binaries host an NS, due to similarities in the SEDs and flux patterns, although there are differences in a case-to-case basis. Even though the emission in gamma-ray binaries is modulated with the orbital period and some even display super-orbital modulation~\citep{Ahnen2016A&A...591A..76A}, the systems LS I +61 303 and HESS J0632+057 have shown enhanced transient episodes ~\citep{2016ApJ...817L...7A,Adams2021ApJ...923..241A}. Since both systems are composed by massive Be stars with a circumstellar disk, the origin of this transient emission is suggested to be associated to clumps or inhomogeneities in the stellar wind or in the interaction region between the stellar and pulsar winds (assuming that HESS J0632$+$057 also host a NS). In the case of long period binaries such as PSR 2032$+$4127 (50~years) and PSR B1259$-$63 (1237 days), their gamma-ray emission is detected during the periastron passage, where VHE signal is detected~\citep{Abeysekara2018ApJ...867L..19A, HESS2020A&A...633A.102H}. In the case of PSR B1259$-$63, some additional HE flares with no flux increase in the TeV counterpart are detected during the periastron passage. 

Microquasars are binary systems composed of a compact object accreting material from a companion star, generating accretion disks and jets. In the HE regime, two microquasars, both of them hosting a massive donor star, have been identified to emit transient emission: Cygnus X-1~\citep{Bulgarelli2010,Sabatini2010,Sabatini2013} and Cygnus X-3~\citep{Tavani2009,Fermi-LAT2009}. No HE emission from microquasars with a low-mass companion (so-called low-mass X-ray binaries, LMXBs) has yet been detected in the MeV range, with the strongest hint being that at $\sim$4$\sigma$ level of V404 Cyg during the major outburst of 2015 \citep{Loh2016,Piano2017}. No transient emission from microquasars has been detected in the TeV domain; see, e.g.,~\citep{scox1,mwc656,Ahnen2017MNRAS.471.1688A,HESS2018mic,maxi2022}. The strongest TeV hint of emission up to now is that of Cygnus X-1 reported by~\citep{Albert2007} during a contemporaneous X-ray flare, reaching a 4.1$\sigma$ (post-trial) signal in an 80 min observation. Only two microquasars have been discovered to emit persistent TeV emission: SS433, a microquasar with two persistent jets that interact with a surrounding nebula, being the interaction regions between the jet and the nebula the TeV-bright region~\citep{SS433_HAWC}, and the microblazar V4641 Sgr~\citep{2023JPhCS2429a2017T}. Both were discovered by particle detectors after accumulating few years of data. Only SS433 has been detected by an IACT~\citep{2024Sci...383..402H} after accumulating more than 200~h of data.

\section{Supernovae}
\label{sec_sne}

SNe are explosive energetic events that result from a stellar death. They can generally be classified into two large groups depending on whether their spectra is hydrogen-poor (type I SNe) or if they do show Balmer lines (type II SNe). More refined classifications have been appearing in terms of additional spectral features, see e.g.,~\citep{Filippenko1997ARA&A..35..309F} or even depending on whether the optical light curve decays (in magnitudes) linearly (II-L) or forming a plateau~(II-P).  

These SNe also differ in the nature and structure of the stellar progenitor. Type Ia SNe result from the thermonuclear runaway of a WD in a binary system. The WD has been accreting material from its companion star and it has probably been producing nova eruptions throughout its life. Due to accretion, the WD increases its mass until the Chandrasekhar limit ($M_{Ch}$ $\approx 1.4 M_{\odot}$) is reached and it will explode as a SNe. As mentioned in Section~\ref{sec_novae}, symbiotic systems such as RS Oph are type Ia SNe progenitors. Core-collapse SNe (CCSNe) are the consequence of the death of a massive star ($M > 8~ M_\odot$) that has exhausted its fuel, producing a violent explosive release of the external shells and leading to the collapse of the nucleus. Type Ib and Ic are \textit{stripped envelope} SNe, in which the progenitors were Wolf Rayet stars stripped of their H (Ib) and He (Ic) layers~\citep{Georgy2009A&A...502..611G}. The progenitors of II-P SNe are generally red-supergiant (RSG) stars and those of II-n are luminous blue variables (LBVs). These LBVs are very massive ($M > 25~M_\odot$) and show high eruptive mass-loss processes during their lifetime, leading to the appearance of strong narrow H emission lines due to the interaction of the SNe with the surrounding circumstellar medium (CSM) in the spectra of II-n. RSGs can also show some smaller eruptive events and hence also show early CSM interaction \textit{flash} features in their early spectra. The progenitors of IIb SNe are supergiants that were partially stripped from their H envelope via binary interaction during the pre-SN phase. 

CCSNe are of interest due to many aspects. They are the precursors of compact objects (BHs and NS, depending of the initial mass of the progenitor star), they help disperse the heavy elements that have been created by the progenitor star onto the interstellar medium (ISM), and they are sources of cosmic rays, neutrinos, and (likely) GWs. Finally, some supernova remnants (SNRs) have been suggested as counterparts of sources detected up to 100 TeV \citep{2021Natur.594...33C,2023arXiv230517030C} which can contribute to the Galactic cosmic-ray spectrum. The most common SNe are type II-P, representing 57$\%$ of the population \citep{Smith2011MNRAS.412.1522S}. 

\subsection{Expected Gamma-Ray Emission from SNe}
SNe are expected gamma-ray production sites due to the acceleration of protons in the SNe blastwave \citep{Murase2011PhRvD..84d3003M, cristofari22,Brose2022MNRAS.516..492B}. They would then contribute to the Galactic cosmic-ray budget. To account for the measured CR spectrum, about 10$\%$ of the ejecta energy ($\sim$10$^{51}$ erg) shall be converted into kinetic energy.

\subsection{Type Ia SNe}

Type Ia SNe are thermonuclear explosions generated by carbon--oxygen WDs in binary systems once they exceed their Chandrasekhar's limit. These SNe are used as standard candles since they display similar lightcurves and homogeneous absolute magnitudes, which are used to estimate cosmological parameters. We now know that novae, which are the progenitors of this type of event, are HE (classical and symbiotic) and VHE (recurrent symbiotic novae) gamma-ray emitters. However, no type Ia SNe has been identified as a gamma-ray source. The only observation of this kind was performed by the MAGIC telescopes of SN 2014J \citep{Ahnen2017A&A...602A..98A}, setting the firsts and only ULs to the VHE gamma-ray emission of these explosive events. SN~2014J was discovered in 21 January 2014 in M~82 at 3.6~Mpc and was observed by MAGIC for about 5.4~h starting 6 days after explosion and over a total of four nights.  The integral flux UL set at 300~GeV is 1.3~$\times$~10$^{-12}$ photons cm$^{-2}$ s$^{-1}$. The expected gamma-ray emission in type Ia SNe should be of hadronic nature (which is in line with the hadronic origin of the VHE signal detected in the nova RS Oph) as described in~\citep{Dwarkadas2013MNRAS.434.3368D}. This model suggests that the hadronic emission shall increase with time (for a constant density medium). The gamma-ray emission should have come from the interaction of the protons accelerated in the SN shock with the surrounding medium. Adopting this model, ref.~\citep{Ahnen2017A&A...602A..98A} calculated that the putative gamma-ray flux shall be at the level of  $\sim$1.3~$\times$~10$^{-24}$ photons cm$^{-2}$ s$^{-1}$, well below the sensitivity of IACTs.

\subsection{Core-Collapse SNe}

CCSNe are considered the best candidates for gamma-ray factories. The interaction of the SN ejecta with the surrounding CSM will produce the~GeV-TeV gamma-ray emission via proton--proton interaction with the ambient matter. Hence, CCSNe with strong CSM interaction (type II-n, Ibn, or near II-P with early CSM interaction) are the best candidates for gamma-ray production. The CSM density decreases with increasing distance; hence, the expected~GeV-TeV emission shall take place during the first days of the SNe explosion. However, eruptive phases of the progenitor star before during the pre-eruption can cause the CSM to be layered in shells. These shells could then enhance the expected gamma-ray radiation at later times once the ejecta reaches them. However, the gamma photons can also interact with the low-energy photons from the photosphere, producing pair production and hence leading to the strong absorption of the~GeV-TeV signal during the first days after explosion \citep{cristofari22,Brose2022MNRAS.516..492B}.

No transient emission from SNe has been confirmed in the HE regime. Two candidate sources have been observed in \textit{Fermi}-LAT data, corresponding to the position of two CCSNe~\citep{Yuan_Fermi,Xi_Fermi}, but due to the large uncertainties in the localization regions which overlap with other gamma-ray sources, their confirmation is challenging. Variable gamma-ray emission has been detected to correspond with the peculiar luminous type II SN iPTF14hls located at 150~Mpc \citep{Yuan_Fermi}. It shows similar spectra to H-rich SNe but with a different lightcurve and it is located at the same position of another eruption detected in 1954, which is difficult to explain in an SN context. Also, there is a blazar inside the error box. The HE source is detected between days 300 and 850 after the explosion. However, if the association between the HE source and SN iPTF14hls is real, it would be the first SN to be detected in the gamma-ray domain. Although it is challenging to explain the gamma-ray emission via particle acceleration in shocks, since the efficiency should be too high. In the case of Type II-P SN 2004dj located in the galaxy NGC 2403 (3.5~Mpc), HE gamma-ray emission from the young SNR has been detected after the accumulation of 11.4 years of data \citep{Xi_Fermi}. The authors suggest that this source, whose emission is banishing over time, is the result of the interaction of the SN ejecta with a high-density shell. Two more candidates have been reported by \citep{Prokhorov2021MNRAS.505.1413P} associated with SN AT2018iwp and SN AT2019bvr, with transient HE signals starting 3 and 6 months after the SN explosion respectively. 

\begin{figure}[H]
    \centering
    \includegraphics[width=6.8cm]{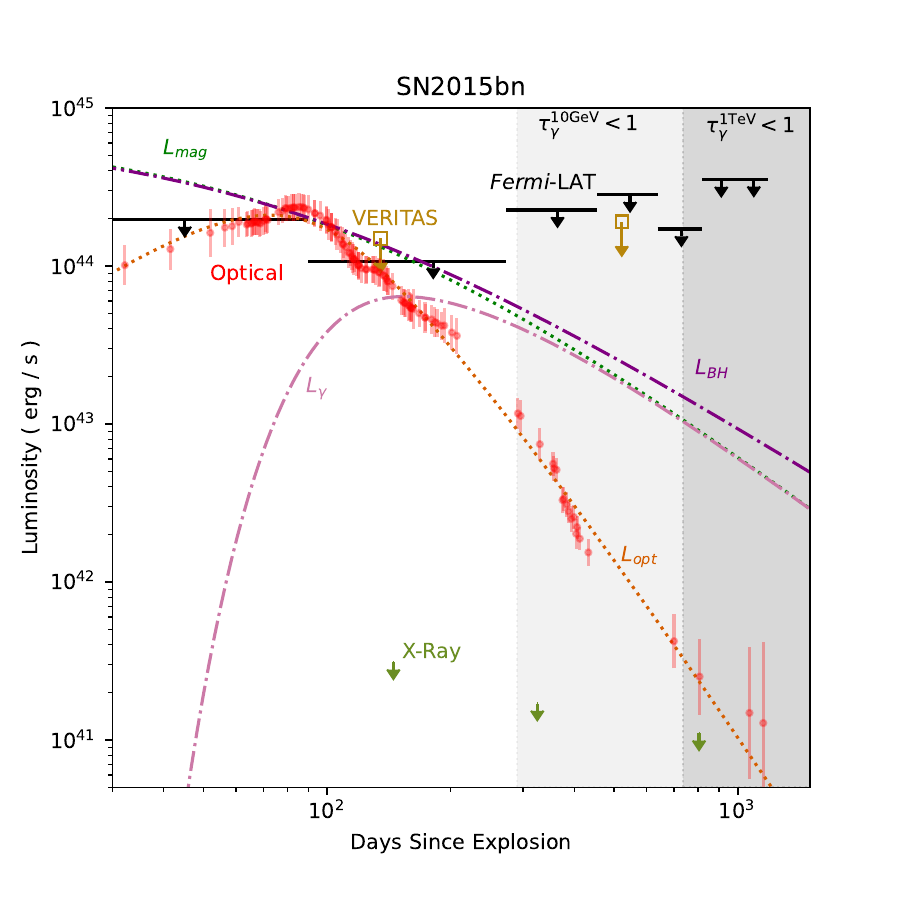}
    \includegraphics[width=6.8cm]{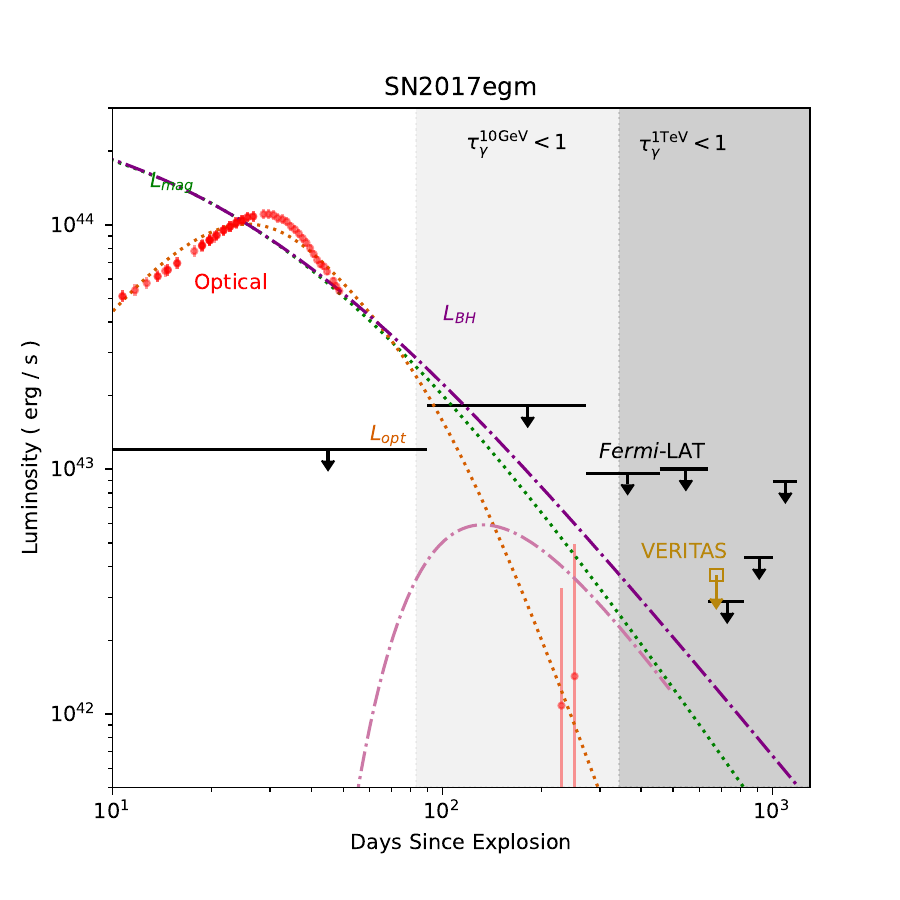}
    \caption{Multi-wavelength lightcurves of SN2015bn from days 30–1500 after explosion (\textbf{left} panel) and SN2017egm spanning 10–1300 days after explosion (\textbf{right} panel). The gamma-ray ULs from VERITAS (orange arrows) and \textit{Fermi}-LAT are shown. Figure adapted and reprinted with permission from \citep{Acharyya2023ApJ...945...30A}. \label{fig_VERITAS_SNe}}

\end{figure}

No signal from CCSNe has been detected in the VHE regime. The authors in  \citep{Abdalla2019A&A...626A..57H} reported ULs on ten different CCSNe observed within a year of the explosion. Nine of them where serendipitously observed, most of them type II-P and located at distances 4--54~Mpc, while ToO observations were performed on SN 2016adj, a type IIb SN located in Cen A galaxy at 3.8~Mpc. The exposure time is also different depending on the target: for four sources including SN 2016adj, observations started around or shortly after the discovery date, while the rest of the observations happened months later (up to 272 days after the explosion). The flux ULs above 1 TeV are of the order of 10$^{-13}$ TeV cm$^{-2}$ s$^{-1}$. The non-detection of this sample may simply indicate that the putative gamma-ray fluxes are below the sensitivity of current IACTs or that the CSM is not dense enough for particle acceleration, but do not rule out the possibility of SNe being VHE emitters. Most recently \citep{Acharyya2023ApJ...945...30A} observed the superluminous (SLSN) type I SN2015bn and SN2017egm. SLSNe are characterized for displaying luminosities 10 to 100 times larger than ordinary CCSNe and for their lightcurves reaching the peak emission at later times. The VHE observations happened 135 days (49 days from the peak magnitude) after explosion for SN2015bn (serendipitously observed) and 670 days from explosion for SN2017egm, targeted due to the predicted gamma-ray emission derived from the optical lightcurve. No TeV counterpart was detected and the first ULs on type I SLSNe in this regime are set (see Figure~\ref{fig_VERITAS_SNe}). Although these ULs do not help constrain the scenarios of a magnetar as central engine or a shock-acceleration they do discard a jet model powered due to fallback accretion onto a black hole (model L$_{BH}$ as seen in Figure~\ref{fig_VERITAS_SNe}). 

The most famous CCSNe is probably SN~1987A, the closest explosion in over 300 yr, located in the Large Magellanic Cloud (LMC), and that reached a peak magnitude of 2.9, visible to the naked eye. It is classified as type-II peculiar, since the progenitor was not a RSG but a blue supergiant. Neutrinos were detected a few hours prior to the arrival of electromagnetic emission, likely happening during the collapse of the progenitor's nucleus~\mbox{\citep{1987PhRvL..58.1490H,1987PhRvL..58.1494B}}. The central compact object is a neutron star powering a pulsar-wind nebula (PWN)~\citep{Cigan2019ApJ...886...51C,Greco2021ApJ...908L..45G}. The evolution of the remnant has been studied over decades in which rings have been formed which are the result of the interaction of the ejecta with pre-eruption material ejected by the progenitor star; see, e.g.,~\citep{McCray2016ARA&A..54...19M}, and references therein for a review. This interaction shall be capable of producing gamma-ray signal via CR acceleration and magnetic field amplification. However, no VHE signal has detected on the remnant after a intensive campaign of 210 h \citep{Abramowski2015Sci...347..406H}.

\begin{figure}[H]
    \includegraphics[width=7cm]{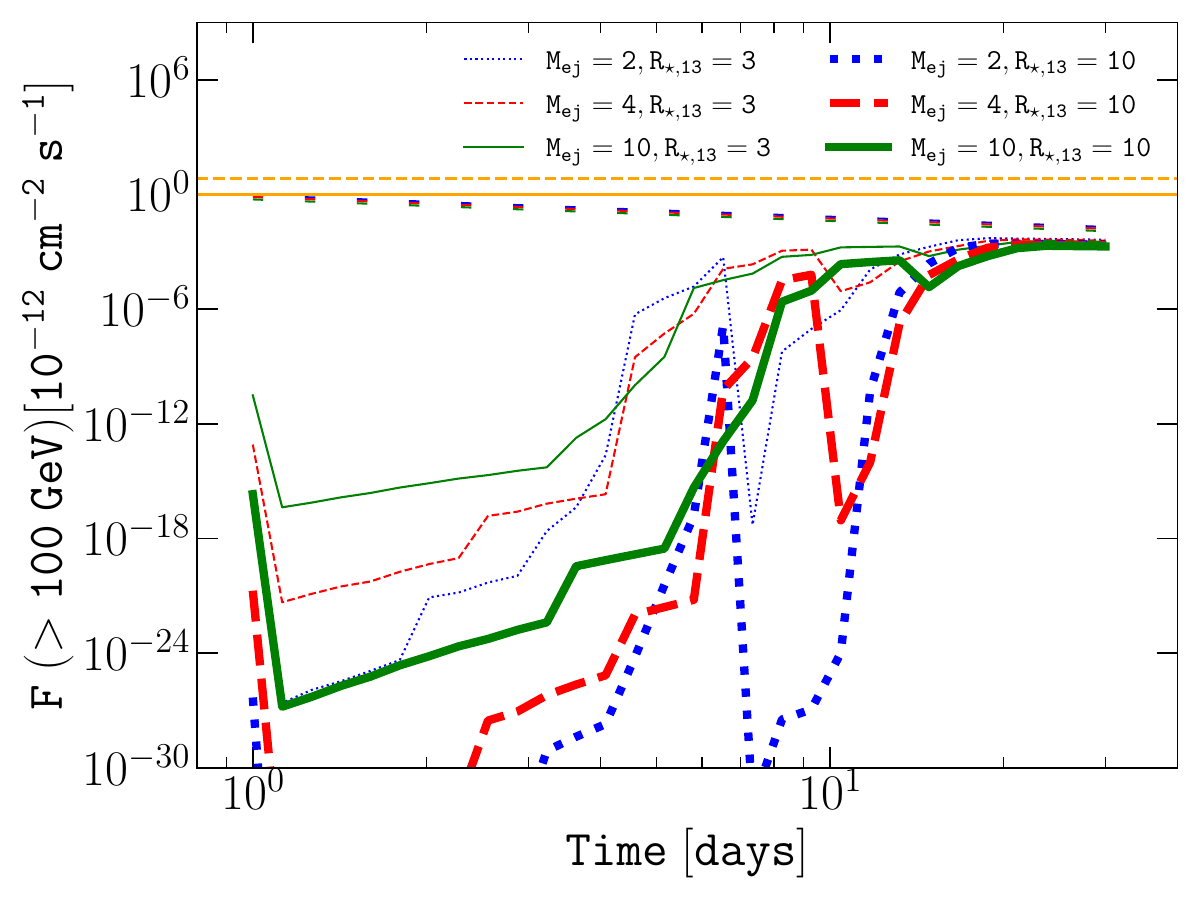}
    \includegraphics[width=7cm]{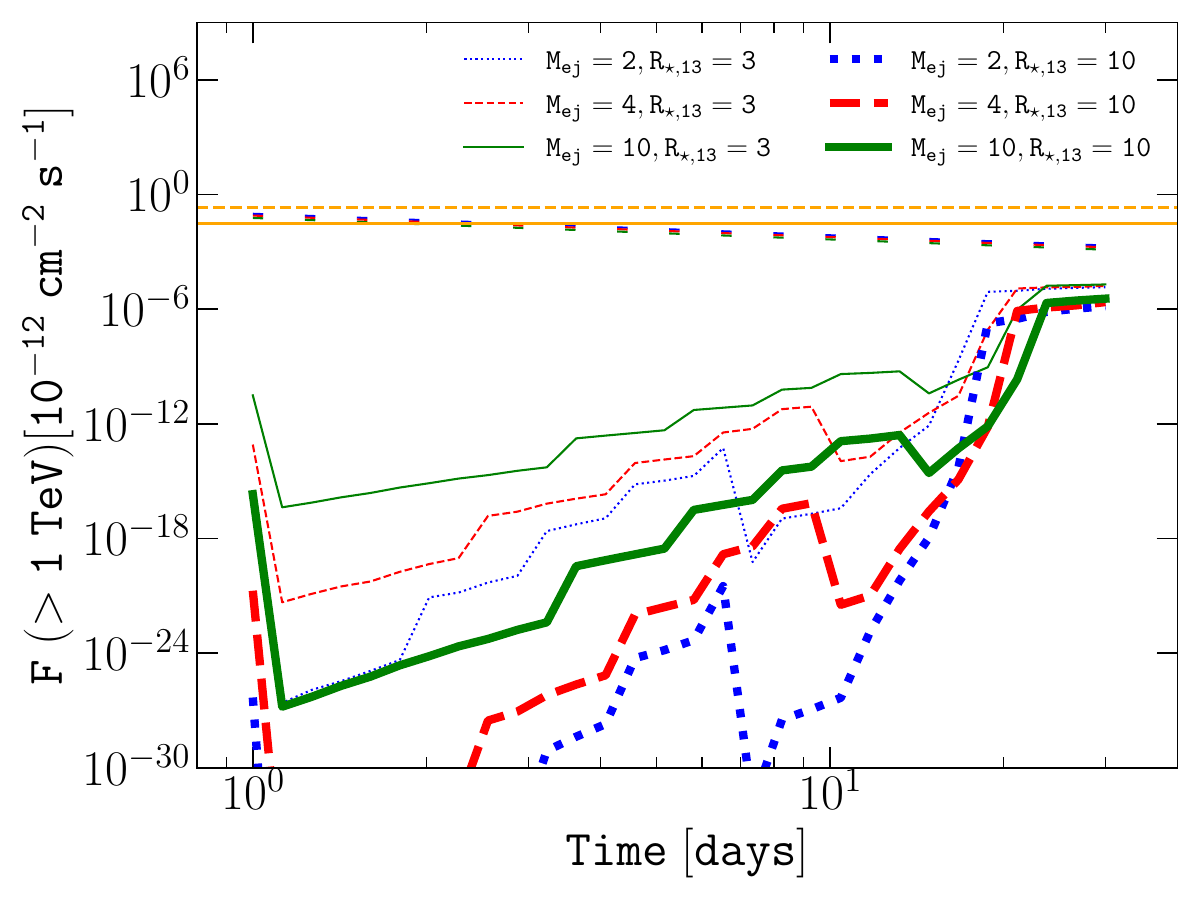}
    \caption{Temporal evolution of the integrated fluxes for type II-P CCSNe at E~$>$~100~GeV (\textbf{left}) and E~$>$~1~TeV (\textbf{right}) at a distance of 1~Mpc and a mass-loss of 10$^{-6}~M_{\odot}$ yr$^{-1}$ for different mass ejecta (blue dotted, red dashed, and green solid) and progenitor radius (thin and thick lines). Figure adapted and reprinted with permission from \citep{cristofari22}. \label{fig4_SNflux}}
\end{figure}

An important parameter to take into account for CCSNe is that the gamma photons can also interact with the low-energy photons from the photosphere producing pair production and hence leading to strong absorption of the~GeV-TeV signal during the first days after explosion. Ref.~\citep{cristofari22} created a time-dependent model to estimate the gamma-ray emission from type II-P CCSNe (the most abundant type of SNe) during the first month after the explosion, taking into account the expected attenuation. By studying the evolution and dependence of different parameters such as photosphere temperature, the authors probe that the expected signal for type II-P CCSNe at distances $>$1~Mpc is below the sensitivity of current generation of IACTs, but close-by systems located in the Galaxy or Magellanic Clouds-located SNe could be detectable now and will undoubtedly be by future-generation of IACTs. Regarding $\gamma\gamma$ effects, the expected signal shall be strongly absorbed during the first 10 days approximately (see Figure~\ref{fig4_SNflux}). Ref.~\citep{Brose2022MNRAS.516..492B} developed a model for Type II-n associated to an LBV progenitor and Type II-P associated to a RSG, accounting also for the strong $\gamma\gamma$ absorption expected during the first days after the explosion. Assuming high mass-loss rates of the progenitor before the eruption, the maximum energies reached by protons can reach up to 600 TeV, which could be compatible with the knee feature of the CR spectrum. However, moderate mass-loss rates show lower values for these energies, between 70 TeV (type II-P) and 200 TeV (type II-n). Considering the absorption effects, the expected gamma-ray peak should happen 12 to 30 days after the explosion. The models suggest that current-generation instrumentation should be able to detect nearby events, up to $\sim$60 kpc for type II-P and $\sim$1~Mpc for type II-n. Future instruments (such as the southern array of the CTA observatory) shall detect type II-P up to 200 kpc and type II-n up to 3~Mpc. The values obtained by \citep{cristofari22,Brose2022MNRAS.516..492B} are in agreement with the lack of detection of a VHE counterpart. 

The second brightest type II CCSNe since the discovery of SN 1987A is SN 2023ixf which went off on May 2023 in M101 galaxy at $\sim$6.8~Mpc and reached a peak magnitude of about B~=~10.6. It is a type II-P SNe with a RSG progenitor that showed strong interaction with the CSM, revealed via flash spectroscopy. No gamma-ray emission has been detected at HE \citep{Marti-Devesa2023ATel16075....1M} and none has been reported at VHE. However, considering the type of explosion and its distance, a non-detection is compatible with respect to current models.  

\section{Flaring Pulsar-Wind Nebulae}
\label{sec_pwne}

Pulsars are highly magnetized neutron stars which are the aftermath of a massive star death. They accelerate leptons (electrons and positrons) in a relativistic wind that halts at the termination shock, creating diffuse structures known as pulsar-wind nebulae (PWNe). In early times, as pulsars are the result of a SN explosion, both the pulsar and its associated nebula are initially surrounded by an SNR. PWNe represent one of the largest VHE source population in the Galaxy. The recent detection (and highly probable association with PWNe counterparts in other cases) of several systems at  $>$100 TeV (ultra high energies, UHEs) has revealed that (at least) a fraction of PWNe are leptonic PeV accelerators or leptonic PeVatrons \citep{2021Natur.594...33C,2023arXiv230517030C}.

PWNe show different evolutionary stages; see, e.g.,~\citep{Gaensler2006ARA&A..44...17G,Giacinti2020A&A...636A.113G,2023Univ....9..402O}: (i) an early phase (typical time t~$\le$~10 kyr) of free--free expansion in which the PWNe are contained inside the SNR and there is no interaction yet with the inward SNR reverse shock. At this early stage, the pulsar is located near the SNR center close to its birthplace. The TeV emission should come, in this case, from within the nebula itself; (ii) the \textit{reverberation} or second phase begins once the reverse shock collides with the PWN forward shock (t$\sim$few tens of years kyr), creating a compression on the PWNe that then leads to an expansion, creating oscillations or reverberations. At this point, the PWN becomes disrupted, provoking that the electrons that produce the TeV emission start to escape from the PWNe onto the SNR and possibly into the ISM. The PWNe start suffering morphological changes and the pulsar can start moving from its birthplace, but it is still contained within the nebula; (iii) the final post-reverberation or bow-shock phase is reached once the pulsar abandons the SNR onto the ISM (at least t~$\ge$~40~kyr), creating a bow-shock structure. At this stage, the escaped leptons can propagate further into the ISM in a region larger than the PWN, creating extended TeV halos. Two TeV halos were initially discovered by HAWC around the evolved pulsars Geminga and Monogem~\citep{2017Sci...358..911A} and several more have been identified by the Large High Altitude Air Shower Observatory (LHAASO) as counterparts for some of the sources on their first catalog~\citep{2023arXiv230517030C}.

The Crab Nebula is probably the most studied PWNe in the VHE regime. It was the first TeV source ever detected \citep{Weekes1989ApJ...342..379W} and since then it has served as standard candle for VHE astronomy.  It is the result of an SN explosion that happened in 1054 CE; hence, it is powered by a young 33-ms pulsar (PSR J0534$+$220). This central pulsar has largely been studied by IACTs and its pulsations have been detected from as low as 25~GeV~\citep{Aliu2008Sci...322.1221A} up to 1.5 TeV energies \citep{Ansoldi2016A&A...585A.133A}. It is the most powerful pulsar in the Galaxy with a spin-down luminosity of 4.6~$\times$~10$^{38}$ erg s$^{-1}$. The multi-wavelength emission of the Crab Nebula is described by synchrotron radiation detected from radio to HE gamma rays, while the TeV component is produced by inverse Compton up-scattering of those synchrotron photons by the relativistic electrons accelerated in the pulsar wind. The TeV PWNe has been resolved to an extension of $\sim$52~arcsec \citep{HESS2020NatAs...4..167H}. Its spectrum is measured over 22 decades in energy, described as leptonic emission. The existence of photons at energies  $>$100 TeV was first evidenced by the UHE detectors Tibet AS$\gamma$ and HAWC  \citep{Amenomori2019PhRvL.123e1101A,Abeysekara2019ApJ...881..134A} and by the MAGIC telescopes via very large zenith angle (VLZA) observation techniques \citep{2020A&A...635A.158M}. LHAASO has further established the Crab Nebula as a leptonic PeVatron with the detection of photons up to 1.1 PeV \citep{2021Natur.594...33C,2023arXiv230517030C}, implying that the parent electrons shall have energies of $\sim$2 PeV.

For a long time, the Crab Nebula was considered a steady source with a stable flux both in the HE and VHE gamma-ray regimes. However, strong flaring emission was discovered at energies  $>$100 MeV by the space-borne instruments AGILE and \textit{Fermi}-LAT~\mbox{\citep{2011Sci...331..736T,2011Sci...331..739A}}. These flux enhancements show a short few-hour timescales duration. The flux increase has been found to vary for a factor 3--30 with respect to the nebula average value, as seen in~\citep{2014RPPh...77f6901B}, and references therein. These flaring episodes can last for several days/weeks and they show shorter-scale structures. Also, the spectral index shows differences among flares. During these flaring events, no variability in the pulsar flux or significant glitch is detected. Also, no variability in the synchrotron component is detected in the radio, infrared of X-ray bands \citep{2011Sci...331..736T,2014RPPh...77f6901B}. These flares have been repeatedly appearing at rates of about $\sim$1--2 per year.

\begin{figure}[H]
\centering
\includegraphics[width=12cm]{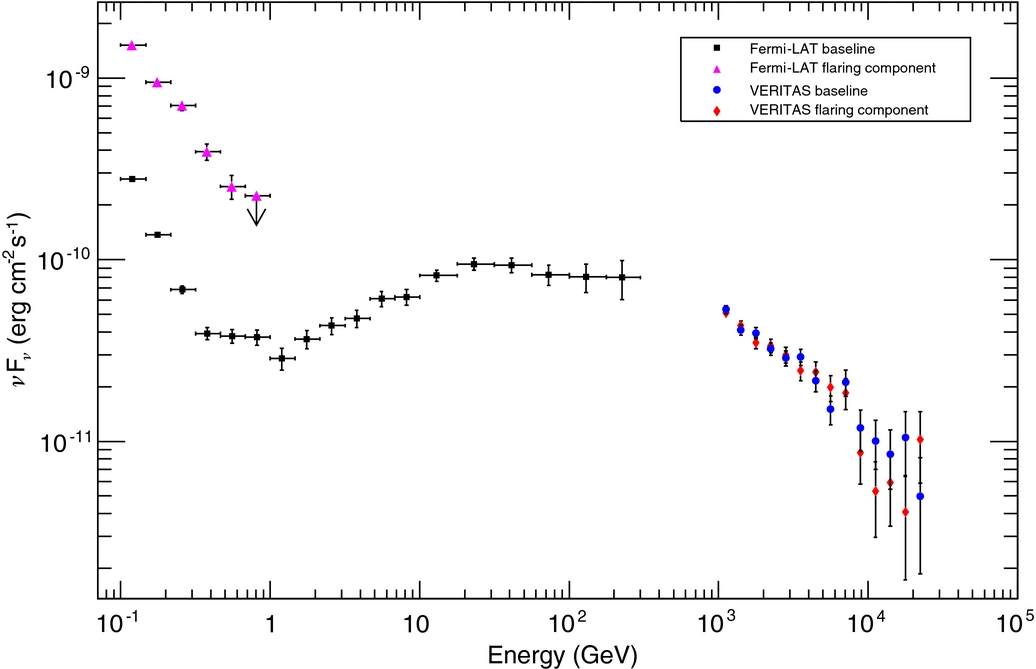}
\caption{SED of the Crab Nebula during quiescence (black squares for \textit{Fermi}-LAT data and blue dots for VERITAS data) and during the 2013 flare (magenta triangles for \textit{Fermi}-LAT and red dots for VERITAS). While the synchrotron component detected by \textit{Fermi}-LAT shows hardening and enhanced flux, the IC spectrum shows no deviation. The y-axis error bars represent the statistical uncertainties in the flux. The downside arrow in the \textit{Fermi}-LAT flaring component is a flux UL.  Reprinted with permission from \citep{Aliu2014ApJ...781L..11A}. 
}.  
\label{fig_veritas_crab}
\end{figure}

This enhanced emission could be extended up to TeV emission and be potentially observable by IACTs. Two scenarios are possible: the detection of the synchrotron tail at the low-end of the VHE regime (up to few tens of~GeV) or that the enhanced emission is transferred to the IC component and detected deep in the TeV range (in Klein--Nishina regime) due to synchrotron self Compton process, since the electrons that produce the enhanced MeV emission should upscatter the photons in the nebula to produce TeV emission.
 However, the IC component seems to remain stable during the HE flares, since IACTs have searched for variability in the TeV component, with no significant flux enhancement detected. Both MAGIC and VERITAS observed during the 2010 HE flare (58 min in one night and 120 min over four nights, respectively), with no VHE enhancement reported~\mbox{\citep{2010ATel.2967....1M,2010ATel.2968....1O}}. The HE flare of 2013 \citep{Ojha2013ATel.4855....1O}, which lasted for about 2 weeks at HE, was observed by H.E.S.S. for five consecutive nights and by VERITAS  during a period of about three weeks (see Figure~\ref{fig_veritas_crab}), with similar results \citep{2014A&A...562L...4H, Aliu2014ApJ...781L..11A}. Whether a flux enhancement deep in the TeV component exists remains yet unknown. A search in the TeV end with the VLZA technique with MAGIC revealed that the stereoscopic system should be able to detect fluctuations 2.25~times brighter that the constant PWNe value \citep{vanScher2019ICRC...36..812V}; hence, given the right conditions, these flares could potentially be detected by the current generation of IACTs. 

It is not trivial to understand the undergoing mechanism of these rapid flares. In the PWNe scenario, the MeV-GeV component is described as synchrotron produced by electrons and positrons in a shocked pulsar wind and hence with energy limited by the synchrotron burn-off (assuming an MHD outflow). The flares surpass this value, hence excluding this ideal scenario. On the other hand, rapid flares cannot be explained in an IC context. Different scenarios have been proposed to account for the origin of these HE gamma-ray flares: (explosive) magnetic reconnection events in a highly magnetized plasma \citep{Lyutikov2018JPlPh..84b6301L} or inductive spikes \citep{Kirk2017PhRvL.119u1101K}, among others. The absence of flux enhancement at other wavelengths could indicate that the HE flares are produced by a single population of electrons. However, it is the lack of multi-wavelength detections and possible correlations that make the study of the origin of this flaring emission challenging. The fast variability and rapid enhancement at HE implies that the emission should come from a compact region in the PWNe of $\sim$10$^{-4}$ pc \citep{2014RPPh...77f6901B}. 

PSR J0534$+$220, powering the Crab Nebula, is the most energetic pulsar in the Galaxy (4.8~$\times$~10$^{38}$ erg s$^{-1}$) and it is among the youngest ones. Up to now, the Crab Nebula is the only PWNe showing variable HE gamma-ray emission in the Milky Way. However, two young pulsars in the LMC show similar spin-down powers to Crab: PSR~J0537$-$6910 (1.5~$\times$~10$^{38}$ erg s$^{-1}$) and PSR~J0540$-$6919 (4.9~$\times$~10$^{38}$ erg s$^{-1}$). Flaring-like activity has been detected with \textit{Fermi}-LAT in different bands: 100--300~MeV, 100--300~MeV, and \mbox{1--10}~GeV~\citep{Nizamov2023MNRAS.520.4456N}. Due to the spatial closeness of the two pulsars, it is not possible to identify which of them is responsible for the flares in the first two energy regimes. However, since the angular resolution improves at higher energies, it has been found that both pulsars flare at~GeV. Gamma-ray flares from these pulsars were already predicted by~\citep{Kirk2017PhRvL.119u1101K}. The detection of flaring gamma-ray emission from other PWNe apart from the Crab Nebula could then indicate that this type of variability (e.g., inductive spikes) is common to young powerful~pulsars.

\section{Fast Radio Bursts and Magnetars}
\label{sec:frbs}
\textls[-5]{Fast Radio Bursts (FRBs) are a relatively newly-discovered Jy-level-class of $\sim$millisecond radio transient events of (mainly) extragalactic origin. At present, since the first discovery of the so-called \textit{Lorimer burst}~\citep{2007Sci...318..777L}, about 1000 FRBs have been detected. However, once taken into account, various factors such as the sky coverage of the different instruments, survey threshold, and selection effects, it is possible to conclude that FRBs occur at an extraordinary rate, up to $\sim$$10^4$ per day distributed over the entire sky. This correspond to a rate of $10^{-3}$~yr$^{-1}$ per galaxy, much larger than the GRB rate. Nonetheless, only few tens of FRBs have been associated to their host galaxy (with kpc precision) and only a handful have been localized with enough accuracy to be associated with specific regions within those galaxies. Although from the observational point of view FRBs are similar to the pulses detected from Galactic radio pulsars, the observed flux density coupled with their extragalactic origin indicates a total emitted isotropical equivalent energy ranging from $\approx$$10^{35}$ to $\approx$$10^{43}$ erg, extremely high compared to the radio pulsar standard, but several orders of magnitude smaller than GRBs;~for a general review, see, e.g.,~\citep{2019ARA&A..57..417C,2019A&ARv..27....4P}.}

\vspace{0.1cm}
Most FRBs are one-off events. However, within the population of FRBs detected so far, around 50 events have been observed to produce multiple bursts, the so-called \textit{repeating FRBs}~\citep{2023ApJ...947...83C}. Although sporadic, the repeating behavior of some FRBs allowed for the first time to perform targeted observations to localize the source using interferometry techniques. The first known repeater, FRB~121102, was associated with a low-metallicity star-forming dwarf galaxy at redshift z~=~0.19~\citep{2017ApJ...834L...7T}, while a persistent and compact ($<$$0.7$~pc) radio source of unclear nature was discovered in association with the FRB direction~\citep{2017Natur.541...58C}. High-resolution optical and infrared observations by the Hubble space telescope and the {\it Spitzer} telescope showed that the galaxy optical emission is dominated by an inner star-forming region whose position is consistent (within uncertainties) with the persistent radio source~\citep{2017ApJ...843L...8B}. Such type of galaxy is also the typical host galaxy for extreme transient events such as GRBs or super-luminous supernovae. While the association of FRBs with cataclysmic events may sound natural and was originally proposed as counterpart of FRBs, the bursts of FRB~121102 have not revealed any signature of an afterglow emission and have been found to repeat at a rate short enough to rule any possible explosive mechanism to power them out. A second localized repeater, FRB~20180916B, shows an apparent $\sim$16-day ($\sim$4 days active followed by 12 days of inactivity) periodicity~\citep{2020Natur.582..351C}. It was found to be located at the edge of a star-forming region within a spiral galaxy, without any persistent counterpart associated. A possible periodicity of $\sim$150 days has been found also for FRB~121102~\citep{2020MNRAS.495.3551R}. Some significant differences between the repeaters and the apparent one--off FRBs have been also reported in the literature~\citep{2021ApJ...923....1P}. In particular, repeater bursts seem to be intrinsically broader in width and narrower in bandwidth. The position of active repeating FRBs seems to be consistent with the one of young extreme objects such as magnetars. Magnetars are isolated NS with an extremely powerful magnetic ﬁeld of the order of $10^{14}$--$10^{15}$~G, about 1000 times stronger than a normal NS. In these objects, the observed persistent electromagnetic radiation is likely powered by the decay in the intense magnetic field. On the other hand, magnetars can also undergo flaring episodes with outbursts on different timescales, detectable in X-rays and radio. These are probably caused by large-scale rearrangements of the surface and/or magnetospheric ﬁeld. Interestingly, magnetars can additionally produce giant ﬂares (GFs), which are among the most energetic (\mbox{$10^{44}$--$10^{47}$ erg s$^{-1}$}) Galactic events. 

\begin{figure}[H]
\centering
\includegraphics[width=13cm]{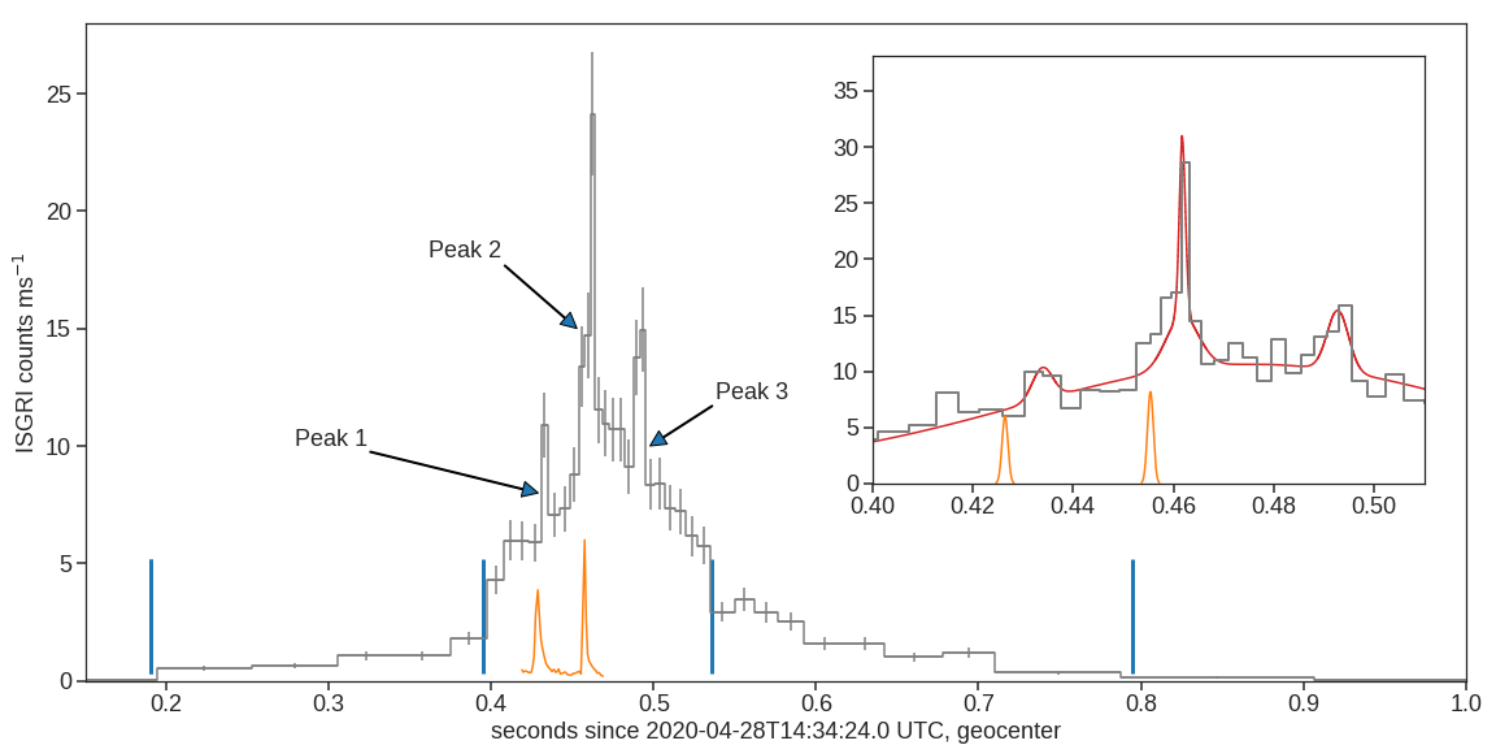}
\caption{INTEGRAL (20--200 keV) light curve of one of the flare of SGR~1935$+$2154 referred to \mbox{T$_0$~=~14:34:24 UTC} of 28 April 2020. The vertical orange lines represent the time of the detected radio pulses. Reprinted with permission from~\citep{2020ApJ...898L..29M}.}
\label{fig:SGR1935}
\end{figure}

\vspace{0.1cm}
On April 2020, the event FRB~200428 was detected by the Canadian Hydrogen Intensity Mapping Experiment (CHIME) telescope from a direction consistent with the Galactic magnetar (and soft gamma repeater) SGR~1935$+$2154~\citep{2020Natur.587...54C}, located at a distance of \mbox{6.6--12.5 kpc} and embedded in the supernova remnant SNR~G57.2$+$0.8. This discovery represented the first detection of an FRB event from a known object, as well as the first FRB of Galactic origin. Contemporaneously to the FRB event, the detection of several X-ray flaring episodes was achieved by a wide range of instruments~\citep{2020ApJ...898L..29M,2021NatAs...5..401T,2021NatAs...5..372R}. Figure~\ref{fig:SGR1935} shows the X-ray light curve as measured by the INTEGRAL satellite where the radio emission is found to be in time coincidence with the X-ray flaring activity~\citep{2020ApJ...898L..29M}. Furthermore, a long-lasting high-energy flaring activity in the form of a forest of intense X-ray bursts was detected by {\it Swift}~\citep{2020ATel13675....1P} and {\it Fermi}-GBM~\citep{2020GCN.27659....1F} up to several hours after the initial episode. The discovery of the connection between hard X-ray bursts (HXRBs) of SGR~1935$+$2154 and FRBs significantly boosted the long-lasting idea of the theoretical interpretation of magnetars as progenitors of FRBs. However, deeper observations performed by the FAST radio telescope\endnote{\url{https://fast.bao.ac.cn/} (accessed on 27 March 2024).} showed that the majority of the X-ray bursts emitted by SGR~1935$+$2154 are actually {\it not} correlated with the FRBs~\citep{2020Natur.587...63L}. Additionally, the further surprisiny detection of the repeater FRB~200120E in a position consistent with a globular cluster within the nearby galaxy M81~\citep{2022Natur.602..585K} challenges the young magnetar scenario as the only engine of FRB. Globular clusters are old enough to not have massive stars able to originate magnetars within. However, they do show high star densities and host short-period binaries which can lead to the production of magnetars via more exotic channels such as accretion-induced collapse (of a WD) or merger-induced collapse (of WDs and NSs). Thus, as a matter of fact, the progenitor of FRB remains a unanswered question. Nonetheless, in light of the so-far only robust hint of association with SGR~1935$+$2154, the magnetar paradigm is still considered the leading interpretative scenario and it has been discussed extensively in the literature~\citep[a non-exhaustive list in ][]{2010vaoa.conf..129P,2014ApJ...797...70K,2016ApJ...826..226K,Zhang2020Natur.587...45Z}.

Within this framework, the proposed emission processes involve coherent radiation mechanisms such as synchrotron maser radiation in magnetar internal~\citep{2020ApJ...896..142B} and external shock models~\citep{2019MNRAS.485.4091M} as well as magnetospheric pulsar-like models. The latter, however, do not foresee keV-MeV emission as observed in SGR~1935$+$2154. In~\citep{2020ApJ...902L..22M}, it is predicted that if FRBs are produced by magnetar flares, an afterglow emission peaking at the MeV-GeV band is expected with a total energy release in the X-ray/gamma-ray band at least a factor $\approx$$10^4$ larger than the emitted radio energy. At the time of writing, the only magnetar flaring event detected in the gamma-ray regime was the GF from a magnetar in NGC~253, detected by \textit{Fermi}-GBM~\citep{GBM2020GCN.27587....1B} and \textit{Fermi}-LAT with a photon of up to 1.7~GeV~\citep{LAT2020GCN.27586....1O}. However, recent results published by {\it Fermi}-LAT on individual FRBs analysis reported no significant emission in the LAT energy band~\cite{2023A&A...675A..99P}. Nonetheless, the detection of hard X-ray bursts with a non-thermal spectrum in SGR~1935$+$2154 shows that at least some FRBs are able to accelerate particles and produce MeV non-thermal emission. Despite the puzzling scenario, FRB (and magnetars) are also an interesting target for IACTs. In fact, some theoretical models also predict VHE~\citep{2014MNRAS.442L...9L,2016MNRAS.461.1498M,2020ApJ...899L..27M,2020ApJ...902L..22M} emission correlated in time with FRBs. Not surprisingly, the flaring episodes of SGR~1935$+$2154 in April 2020 were also followed-up by current Cherenkov telescopes and monitoring campaigns on this magnetar have been active since then. The campaigns organized by MAGIC~\citep{Lopez-Oramas:2021zd} and H.E.S.S.~\citep{2021ApJ...919..106A}, coordinated within a larger multi-wavelength framework, did not reveal VHE emission to be neither persistent nor on shorter (minutes to milliseconds) time scales.

\begin{figure}[H]
\centering
\includegraphics[width=13cm]{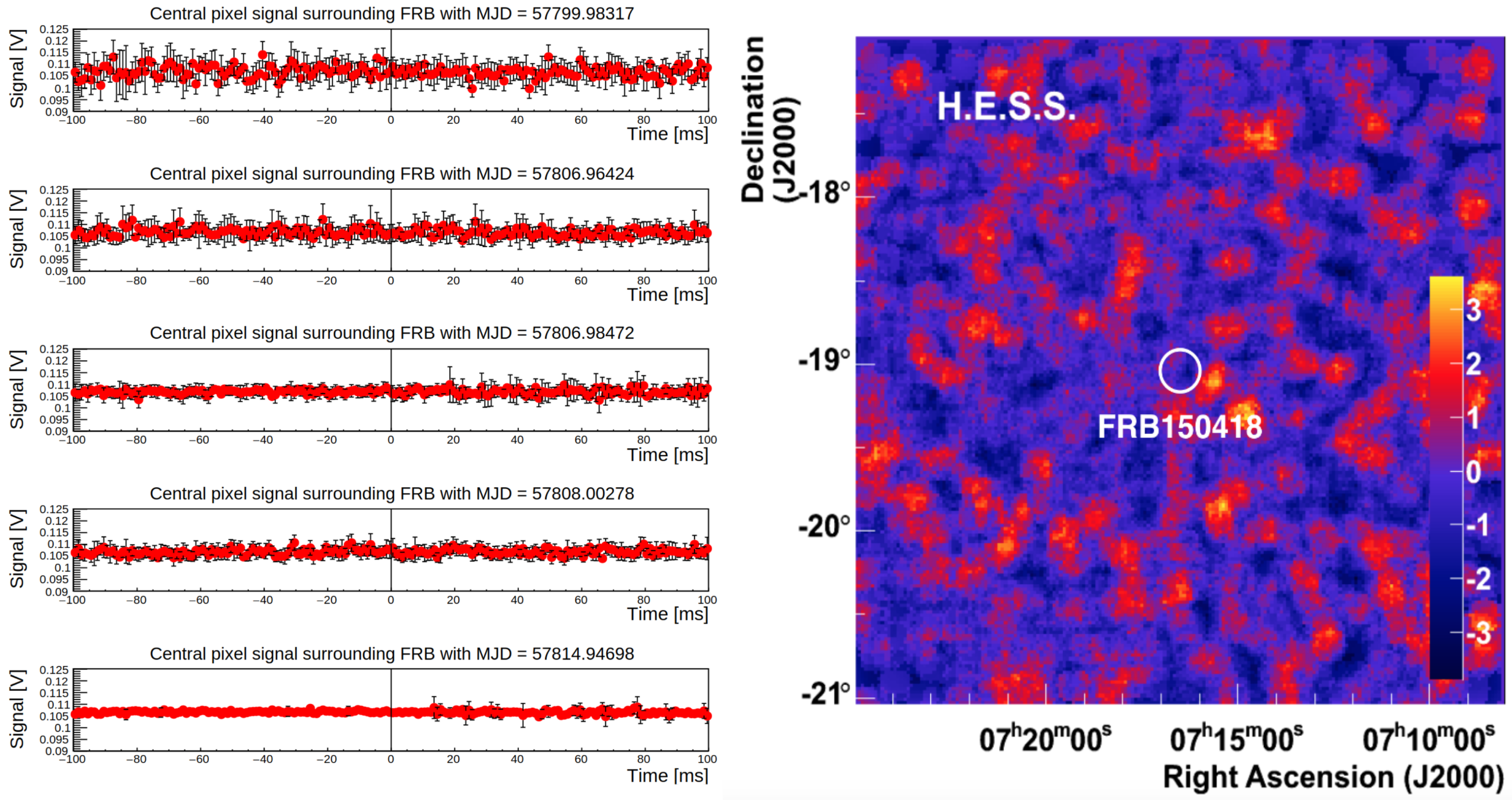}
\caption{({\bf Left} plot): Optical light curves obtained by MAGIC and spanning 200 ms around the trigger times of 5 bursts from FRB~121102 detected simultaneously with MAGIC data. The vertical axis is proportional to the U-band flux. No significant excess is observed simultaneously with any of the 5 bursts. Reprinted with permission from~\citep{2018MNRAS.481.2479M}. ({\bf Right} plot): Significance sky map from the H.E.S.S. follow-up observations of FRB~150418. Reprinted with permission from~\citep{2017A&A...597A.115H}.}
\label{fig:FRBexample}
\end{figure}

\vspace{0.1cm}
Current-generation IACTs have active follow-up programs on FRBs although no particularly stringent flux ULs in the VHE band were reported so far~\citep{2019ICRC...36..698H}. Some specific events have been the subjects of dedicated follow-up campaigns such as in the case of H.E.S.S. that obtained the first ULs on the potential VHE afterglow emission from FRB~20150418A~\citep{2017A&A...597A.115H} (Figure~\ref{fig:FRBexample} right panel). This FRB was of particular interest as it showed hint of a $\approx$$6$~days-long-lasting radio afterglow detected by the Australia Telescope Compact Array (ATCA)~\citep{2016Natur.530..453K}. The achieved limit on the VHE luminosity was of the order of $5 \times 10^{47}$ erg s$^{-1}$ at the energy of 1 TeV. The MAGIC and VERITAS collaborations reported VHE ULs on the repeater FRB~20121102A conducting coordinated observations with Arecibo~\citep{2018MNRAS.481.2479M,2017ICRC...35..621B}. In the case of MAGIC follow-up, five contemporaneous radio bursts were detected (at a central frequency of 1.38 GHz) although no millisecond timescale burst emission was detected in VHE or the optical band (Figure~\ref{fig:FRBexample} left panel). Follow-up results on a sample of other repeaters (FRB~180814, FRB~180916, FRB~181030, and FRB~190116) were more recently reported by VERITAS~\citep{2022icrc.confE.857V}, again with no detection~achieved.

\vspace{0.1cm}
It is important to remark that IACTs are versatile instruments that, although designed to detect nanosecond pulses of Cherenkov light, are generally sensitive to millisecond timescale optical signals. Despite the modest quality of their mirrors when compared to standard optical telescopes, the typical large diameter of their primary mirror and their fast-response readout electronic make them effective high-time-resolution photometers. Current IACTs are indeed able to perform parallel VHE and optical observation on very short timescales up to a limiting magnitude significantly lower than standard optical telescopes~\citep{2020JATIS...6c6002H,2011ICRC....9...38G,2003ICRC....5.2987F}. This is a key feature that make IACTs excellent instruments for fast transient astronomy and with a relevant application in the case of FRBs. In fact, FRBs may be potentially accompanied by \textit{fast optical bursts} (FOBs) via different mechanisms~\citep{2019ApJ...878...89Y}. Optical counterparts have been detected in association with magnetars flaring episodes~\citep{2008Natur.455..503S} and can therefore provide important insights into the physics of FRBs. The possibility to have parallel VHE and fast optical observations made IACTs key instruments for future follow-up. The improvement in VHE sensitivity as promised by next-generation instruments such as the CTA will finally allow for observations up to a gamma emission values comparable to the ones expected by magnetars.


\section{Gravitational Waves}
\label{sec_gws}

\textls[-15]{The possibility of performing astrophysical observations by means of non-electromagnetic signals such as gravitational waves (GWs) has become reality with the first scientific runs of the LIGO\endnote{\url{https://www.ligo.org/} (accessed on 27 March 2024).} and Virgo Scientific Collaborations\endnote{\url{https://www.virgo-gw.eu/} (accessed on 27 March 2024).}~(LVC). The first LVC scientific observation run, named O1, opened the era of gravitational wave astronomy by means of the first direct detection of a GW signal~\citep{2016PhRvL.116f1102A} from a binary stellar-mass black hole merger (BBH). Not long after, during the O2 scientific run, the first GW signal (GW~170817) from the coalescence of a binary system composed of two NSs (BNS) was discovered~\citep{2017PhRvL.119p1101A} together with a new sample of BBH signals~\citep{2019PhRvX...9c1040A}. Approximately 2~s after the detection of GW~170817, the Fermi and INTEGRAL satellites detected a sub-threshold short GRB (sGRB), namely GRB~170817A~\citep{2017ApJ...848L..14G,2017ApJ...848L..15S}. The identification of a sGRB as electromagnetic counterpart of a GW signal triggered by a binary neutron star merger represented a groundbreaking observation that provided the first firm evidence on the nature of sGRB's progenitors. The potential link between GWs (from BNS mergers) and sGRBs has been widely explored and discussed in the literature in the past;~see, e.g.,~\citep{2016SSRv..202...33L}, and references therein for a review. The discovery of GRB~170817A triggered an unprecedented follow-up campaign at all wavebands. It is important to remark that these observations are particularly challenging due to the very large localization uncertainties provided by GWs interferometers, up to tens of thousands of square degrees. Nonetheless, approximately 11~h after the GW trigger, an optical/IR counterpart, named AT~2017gfo (IAU naming) and hosted in the 40~Mpc-distant galaxy NGC~4993, was detected by the One-Meter Two-Hemisphere (1M2H) collaboration~\citep{2017Sci...358.1556C} and interpreted as a {\it kilonova}. Unlike BBH mergers, BNS mergers are expected to be source of optical/near-IR emission powered by the decay of radioactive nuclei generated by r-process nucleosynthesis in the outflow formed after the coalescence;~see, e.g.,~\citep{1998ApJ...507L..59L,2010MNRAS.406.2650M}. The detection of AT~2017gfo represented the first confirmation of this theoretical prediction. In the days after the burst, an X-ray counterpart was detected and identified as the GRB afterglow non-thermal emission~\citep{2017Natur.551...71T}. The late-time rising of an X-ray afterglow fits within the interpretation that the GRB is observed off-axis, with the jet-beamed ejecta pointing away from Earth. According to the hydrodynamic of a generic relativistic shock-wave model, the bulk Lorentz factor ($\Gamma(t)$) of the outflow is reduced by the deceleration of the jet, causing the relativistic beaming angle ($\propto 1/\Gamma(t)$) to increase~\citep{1976PhFl...19.1130B}. The opening angle of the emission widens, eventually including the line of sight of the observer. From the observational point of view, a delayed emission, whose intensity and delay depends on the off-axis angle, may arise. The case of GW/GRB~170817A confirms this paradigm as successive radio observations did confirm the interpretation of the radio to X-ray emission as originated by an off-axis, structured jet (i.e., the energy and velocity of the ejected material scale with the angular distance from the jet axis) with a viewing angle of $\approx$$30^{\circ}$~\citep{2017Natur.551...71T,2017Natur.551...67P,2019Sci...363..968G}. The radio and X-ray emission increased in the weeks following the initial trigger, peaking approximately 155 days after the merger.}

\begin{figure}[H]
\centering
\includegraphics[width=13cm]{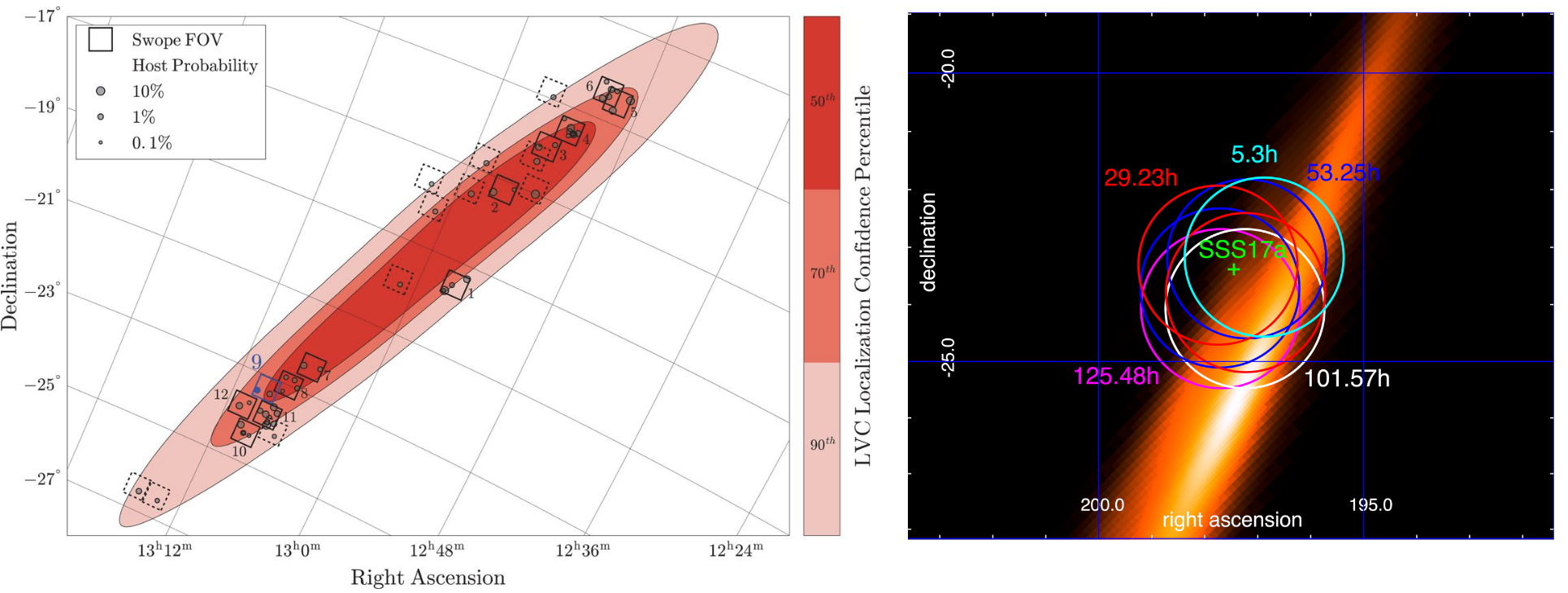}
\caption{({\bf Left} plot): Sky map covering the 90\% confidence-level region for the GW~170817 position. The positions of galaxies used in cross-correlating the large localization area and defining an optimized pointing strategy for the Swope telescope on 17--18 August 2017 are shown as gray circles. The size of the circle indicates the probability of a particular galaxy being the host galaxy for GW~170817. The square regions are individual Swope pointings labeled in the order that they were observed. Solid and dashed squares represent the square chosen to contain multiple and individual galaxies, respectively. The position of NGC~4993 and SSS17a are in the blue square. Reprinted with permission from~\citep{2017Sci...358.1556C}. ({\bf Right} plot): H.E.S.S. pointing directions during the monitoring campaign of SSS17a. The circles denote an FoV with radius of $1.5^{\circ}$ and the shown times are the start times of each observation with respect to GW~170817A. Colored background is the GW localization map. Reprinted with permission from~\citep{2017ApJ...850L..22A}.}
\label{fig:gwpointings}
\end{figure} 

\begin{figure}[H]
\centering
\includegraphics[width=14cm]{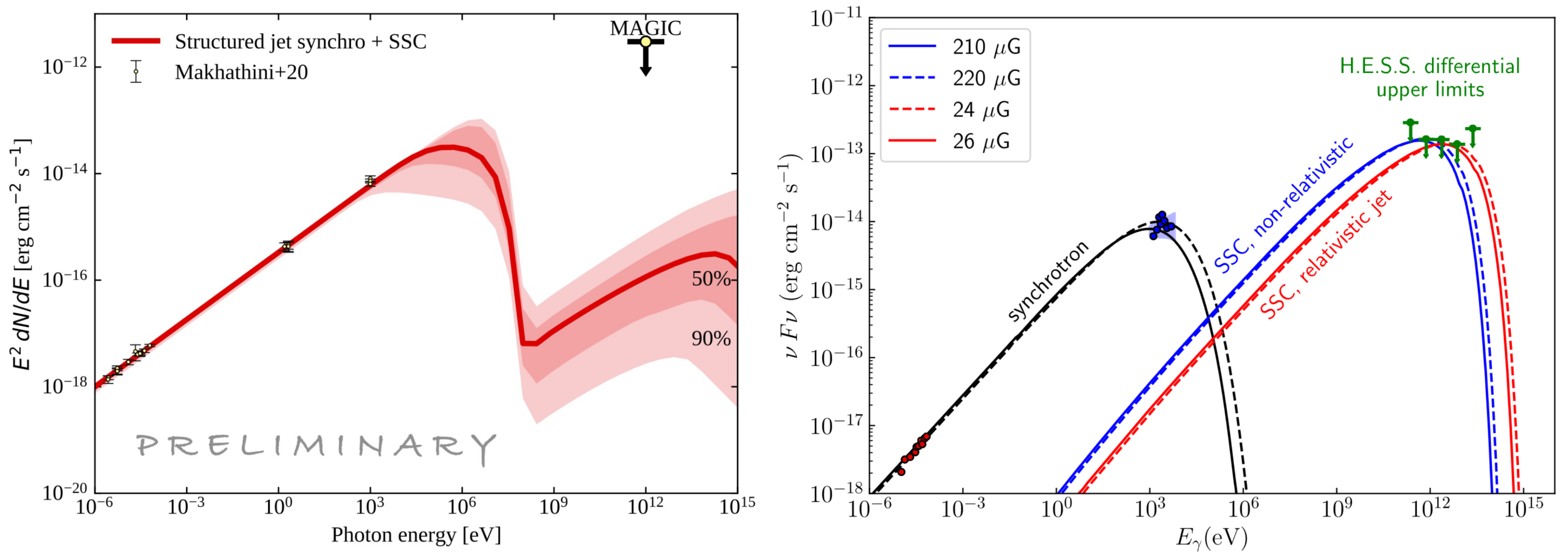}
\caption{({\bf Left} plot): Expected SSC emission evaluated by MAGIC 155 days after the merger, using fit parameters constrained by the radio, optical, and X-ray data. Reprinted with permission from~\citep{2022icrc.confE.944S}. ({\bf Right} plot): SSC spectra foreseen in~\citep{2020ApJ...894L..16A} 110 days after the merger. The blue and red curves represent two possible geometry and expansion speed of the remnant: an isotropic, non-relativistic expansion (blue curves) and a relativistic jet (red SSC curves). The minimum magnetic field strength imposed by the H.E.S.S. ULs (green arrows) are also reported. Reprinted with permission from~\citep{2022icrc.confE.944S}. }
\label{fig:gwSEDmodels}
\end{figure}

The extensive multi-wavelength campaign triggered by the detection of GW~170817A also included follow-up at VHE by IACTs (see Figure~\ref{fig:gwpointings}). Less than two years after GW~170817A, the first detection of VHE gamma-ray emission from GRB~180720B~\citep{2019Natur.575..464A}, GRB~190114C~\citep{2019Natur.575..455M}, and GRB~190829A~\citep{2021Sci...372.1081H} was announced by the H.E.S.S. and the MAGIC collaborations, bringing an end to a quest lasting for more than twenty years. Although all of the GRBs detected so far\endnote{More GRBs have been detected at VHE since then such as GRB~201216C by MAGIC and the remarkable GRB~221009A, although not detected by IACTs, observed by LHAASO up to 13 TeV.} by current IACTs were long GRBs, sGRBs are also expected to emit VHE radiation. In this regard, a hint of VHE emission has been observed by MAGIC in the case of the short GRB~160821B~\citep{2021ApJ...908...90A}, providing a compelling clue on the detectability of TeV emission from GW counterparts from compact object mergers. Few attempts by IACTs in following-up GW alerts were reported before the breakthrough of GW~170817A such as for GW~151226~\citep{2015GCN.18776....1A,2017IAUS..324..287D}, GW~170104~\citep{2017GCN.21153....1M}, and GW~170814~\citep{2019MmSAI..90...49A}. However, the VHE campaign organized for GW~170817A represented a step forward and a fundamental test-bench in exploring IACTs' capabilities in this challenging observations. The H.E.S.S. telescopes started a series of pointing over the uncertainty region of GW~170817A about 5~h after the first trigger~\citep{2017ApJ...850L..22A}, that made it the first ground telescope to point at the source location. Although the detection of AT~2017gfo was not yet announced~\citep{2017GCN.21529....1C}, the pointing strategy proved to be efficient with the NGC~4993 location within the field of view of the H.E.S.S. first pointing. Nonetheless, no evidence of VHE emission was detected during this early monitoring campaign of SSS17a~\citep{2017ApJ...850L..22A}. Starting from mid-December 2017, the sky position of the optical transient SSS17a became visible also to the MAGIC telescopes' site. This observation window roughly overlaps with the afterglow peaking time. Late time follow-up was then performed by MAGIC and, again H.E.S.S. covering the peak and the onset of the fading phase in the X-ray and radio lightcurves. Although no detection was achieved, the obtained ULs were used by the two collaborations to constrain physical emission models, although with a rather different prediction on the intensity of the TeV component (see Figure~\ref{fig:gwSEDmodels}). 

Both in~\citep{2020ApJ...894L..16A,2022icrc.confE.944S}, the broad-band SED is modeled by means of a synchrotron + synchrotron self-Compton (SSC) processes. However, the two proposed models are not directly comparable. In the structured jet approach used within the MAGIC interpretation and described in details in~\citep{2019A&A...628A..18S}, the full time-evolution of the jet is taken into account in the evaluation of the expected emission.
 As the jet evolves, the observed radiation is the result of the convolutions of photons emitted at different times and different locations behind the shock. Such an evolution is not considered within the H.E.S.S. modeling wherein the emission is evaluated at specific single times. Within the uncertainties of assumed physical parameters for GRB~170817A, the structured jet model foresees a rather low TeV emission component, significantly lower than MAGIC upper limits, challenging the possibility of detection of such an event with current-generation IACTs. However, more favorable conditions in the emitting geometry and circumbusrt properties may mitigate these prospects as we will discuss in Section~\ref{sec_discussion}.

\section{Other Transient Sources: Tidal Disruption Events and Gamma-Ray Bursts} 
\label{sec_other}
The science topics of GRBs and tidal disruption events (TDEs) are discussed in more details in dedicated papers within this Special Issue~\cite{2024VL}. Nevertheless, we briefly touch upon them in the following subsections for the completeness of this review.

\subsection{Tidal Disruption Events and Neutrino Connection}
\label{sec:tde}
Tidal disruptions events (TDEs) are powerful events that occur when a star is disrupted by tidal forces when approaching a massive BH. They are considered of extreme importance in particular in the framework of multi-messenger astrophysics. It is thought that the disruption of a stellar object may trigger the launch of a relativistic jet able to shock-accelerate particles from the star remnants. This material is naturally rich in light and heavy nuclei so that TDE may be a plausible acceleration site for ultra-high-energy cosmic rays (UHECR~$>10^{20}$~eV) and neutrino;~see, e.g.,~\citep{2014arXiv1411.0704F,2020A&A...636C...3G}. While the cosmic neutrino flux has already been established through the measurements of the IceCube Neutrino Observatory~\citep{2015PhRvL.115h1102A}, the association of this flux with specific astrophysical sources is still challenging. So far, few sources have been correlated with neutrinos: the flaring blazar TXS~0506$+$056, identified as the potential source of the IceCube neutrino alert IC~170922A~\citep{2018Sci...361.1378I}; the nearby star-forming galaxy NGC~1068, in spatial coincidence with a cluster of IceCube-detected neutrinos~\citep{2022Sci...378..538I}; and the TDE AT~2019dsg, discovered in the optical band by the Zwicky Transient Facility (ZTF\endnote{\url{https://www.ztf.caltech.edu/} (accessed on 27 March 2024).}~\citep{2019PASP..131a8002B}) and identified as the source of the event IC~191001A~\citep{2021NatAs...5..510S}. While NGC~1068 and TXS~0506$+$056 have already been the target of observations with IACTs~\citep{2022ApJ...927..197A,2019ApJ...883..135A}, TDE is a relatively unexplored class of sources in the VHE band. In 2011, the remarkable TDE event Sw~J1644$+$57, originally triggered as a GRB by \textit{Swift}, was the subject of an extensive follow-up campaign by MAGIC~\citep{2013A&A...552A.112A} and VERITAS~\citep{2011ApJ...738L..30A}. Although no significant VHE detection was found, these observations may potentially pave the road to future follow-ups with both current IACTs and the CTA~Observatory.   

In the near future, the Vera Rubin Observatory~\citep{2019ApJ...873..111I}, will start operation and overlap with the CTA Observatory era. The Legacy Survey of Space and Time (LSST) will also dramatically enlarge the sample of detected TDEs, thereby providing an unprecedented number of possible triggers to CTA that may be able to detect VHE gamma-signature at least for nearby events ($\lesssim$$20$~Mpc)~\citep{2016MNRAS.458.3314C}.   

\subsection{Gamma-Ray Bursts}
GRBs are transient events last from milliseconds up to hundreds of seconds. They are the brightest electromagnetic events known and they are able to release an enormous amount of energy ($10^{52} \div 10^{54}$~erg). They show their phenomenology mainly in the \mbox{10 keV--1 MeV} energy band. According to a relativistic shock model, described for example in~\mbox{\citep{1986ApJ...308L..43P,1999PhR...314..575P}}, GRB emission is powered by the conversion of the kinetic energy of a relativistic outflow into electromagnetic emission. The details of this conversion remain poorly understood. A largely discussed possibility is that the observed photons are radiation from particles accelerated to ultra-relativistic energies by successive collisions within the magnetized medium. During the so-called {\it prompt phase}, GRB dynamic is thought to be driven by relativistic collisions between shells of plasma emitted by a central engine (internal shocks). Similarly, the emission during the {\it afterglow} phase seems to be connected to the shocks between these ejecta with the external medium (external shocks). The results of such internal/external shocks is the acceleration of particles through Fermi mechanisms. The accelerated particles can emit the observed high-energy photons through many possible non-thermal mechanisms. Within this framework, synchrotron emission has largely been considered as the most natural to explain the GRB sub-MeV emission~\citep{2001ApJ...548..787S,2001ApJ...559..110Z,2007MNRAS.380...78G}. Although alone it cannot fully explain the observed prompt spectrum for the majority of the events, synchrotron is believed to play an essential role in GRB dynamic. In particular, it has been suggested that the HE emission observed by {\it Fermi}-LAT extending after the end of the prompt emission is synchrotron radiation produced in the external shock that is driven by the jet into the circum-burst medium~\citep{2010MNRAS.403..926G}. However, the recent detection of a VHE counterpart challenged the synchrotron-alone scenario, confirming the existence of a second emission component above the synchrotron burnoff limit. In the near future, the CTA will open the possibility of detecting $\sim$hundreds (or more) of photons from moderate-to-bright GRB, allowing for a significant improvement in the photon statistics and for the possibility to have good-quality time-resolved spectra. The first prototype of the 23 m class diameter LST-1, particularly suited for the follow-up of transient events due to the fast repositioning ($\sim$30 s for 180$^{\circ}$) and the relatively low energy threshold, is currently ending its commissioning phase at the CTA northern site. LST-1 already reported the follow-up of different GRBs and neutrino events, although with no reported significance yet~\citep{2021arXiv210804309C}.   

Furthermore, many events have shown (somehow surprisingly) that long-lasting TeV signatures can also be detectable under favorable conditions. The close-by and very low luminosity burst GRB~190829A~\cite{2021Sci...372.1081H} was detected by the H.E.S.S. telescope up to a few TeV for three consecutive nights while the recent detection of GRB~221009A by the LHAASO experiment up to 13 TeV~\cite{2023Sci...380.1390L} has definitively proven that instruments operating in the energy range above few TeV band such as the ASTRI Mini-Array~\citep{2022JHEAp..35...52S}, although not specifically designed for transients and time-domain astrophysics, may also play a key role in the future follow-up programs of these events~\citep{Stamerra:2021rK,Carosi:2023+W}.
For more detailed review on GRBs, see~\citep{2022Galax..10...67B,2024VL}


\section{Discussion and Prospects}
\label{sec_discussion}

\begin{figure}[H]
\centering
\includegraphics[width=10cm]{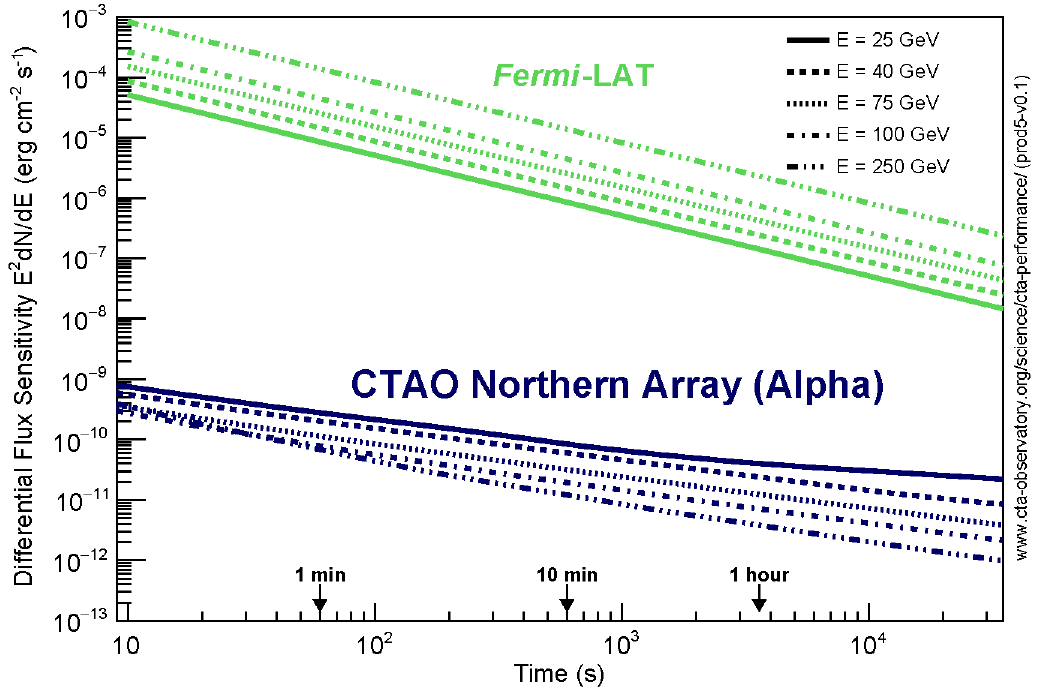}
\caption{Differential flux sensitivity versus time of the CTA Observatory Northern Array in its Alpha configuration (blue), which accounts for the initial construction phase, compared to \textit{Fermi}-LAT (green) at different energies. Figure taken from \url{https://www.cta-observatory.org/science/ctao-performance/} (accessed on 27 March 2024).}
\label{fig:CTAsens}
\end{figure}

The last two decades have proven to be the starting point of a golden era for multi-messenger time domain astrophysics. New facilities for non-electromagnetic astronomy such as neutrino and GW detectors have reached their nominal operational phase, joining a large network of telescopes and satellites covering an unprecedentedly wide energy band. New synergies and improving communication channels between these facilities have led to breakthrough discoveries such as the connection between sGRBs and GWs. The physics of extreme transient events both inside and outside our Galaxy has an intuitive connection with the highest energetic X- and gamma-ray radiation. Non-thermal emission processes, typical of the HE and VHE band, represent the signature of shock-powered radiation mechanisms, often invoked in explaining the dynamics of a wide range of extreme cosmic accelerators. Shock interactions may be at work as a particle acceleration mechanism in both a relativistic (like in GWs/GRBs) and non-relativistic (like in SNe) flavor. The corresponding radiation mechanisms at work may be shared among these sources, although showing a diverse phenomenology given the differences in shocks expanding velocity, external density, and surrounding environment. Hence, VHE observations provide a privileged channel to shed light into the physics of transient events in an energy range particularly important for the discrimination among different emitting scenarios. Although in operation since the beginning of the 2000s, current-generation IACTs can still lead to the discovery of new transient phenomena and to deepen our understanding of the TeV physics of these newly identified VHE sources. However, the IACT community is working toward the construction of the CTA Observatory, which is the next-generation ground-based observatory for Cherenkov astronomy. It will be composed by two arrays composed of telescopes of up to three different sizes, located in the northern (Roque de Los Muchachos Observatory, Spain) and southern (Paranal Observatory, Chile) hemispheres. It will cover the energy range between 20~GeV and 300 TeV and it will count with improved sensitivity with respect to current IACT experiments. It will have unprecedented sensitivity at short timescales (see Figure~\ref{fig:CTAsens}), making it a unique laboratory for VHE transient astrophysics. The \textit{Transients Key Science Project} of the CTA Observatory \citep{CTA2019scta.book.....C} defines the core program for the follow-up of transient sources~\citep{astronet2021arXiv210603621B,Carosi2022icrc.confE.736C,ALO2023hsa..conf..159L}, including GRBs, GWs, neutrino counterparts or the large zoo of Galactic transient sources (novae, microquasars, magnetars, flaring PWNe, etc.), among other serendipitous transitory events.

The improved sensitivity of CTA, together with its better angular and energy resolution and large energy coverage, will allow for the discovery of new transient and multi-messenger phenomena, widening the population of current known sources.

\subsection{Novae}
\begin{figure}[H]
\includegraphics[width=14cm]{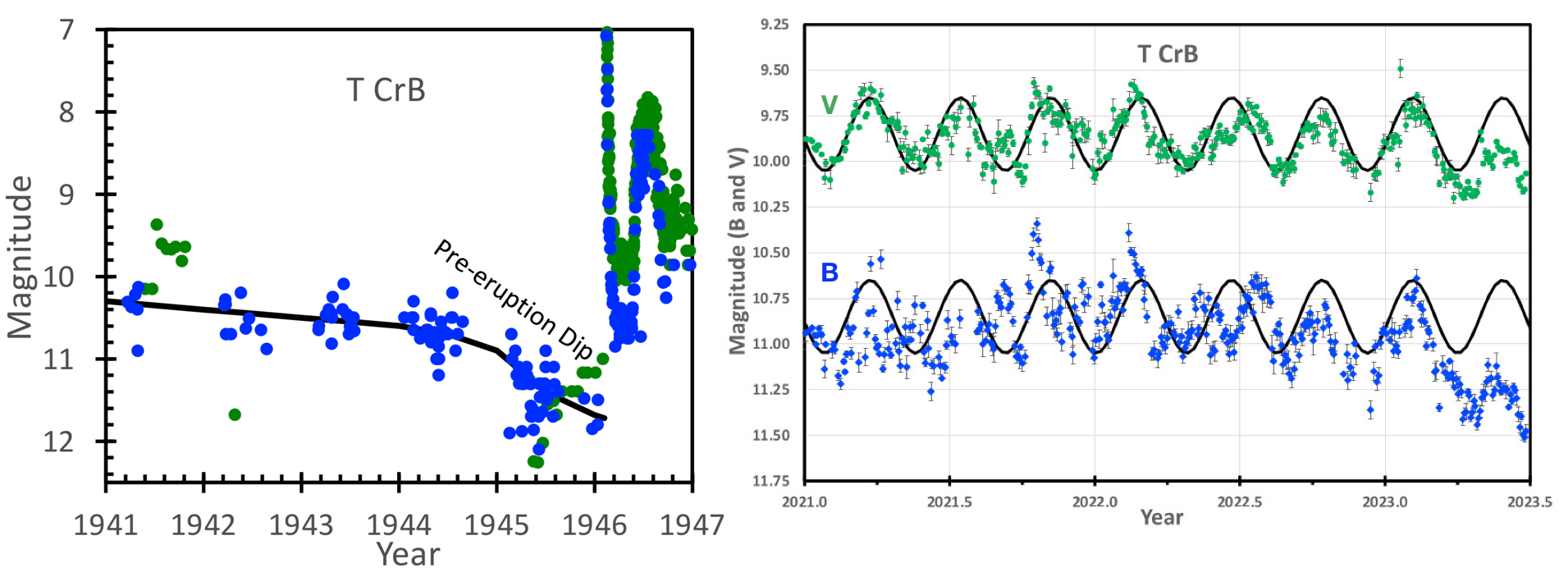}
\caption{(\textbf{Left}) Optical lightcurve of T CrB during the 1946 eruption in B (blue dots) and V (green circles) magnitudes. A similar lightcurve is expected in the next eruptive event; (\textbf{Right}) Current fading of the B (blue circles) and V (green circles) of T CrB, revealing that the source has entered the pre-eruption dip. Figures from \citep{2023ATel16107....1S_TCrB,2023MNRAS.524.3146Schaefer_TCrB}.\label{fig_TCrB}}
\end{figure}

 We can expect to detect other recurrent symbiotic systems in the VHE regime. There are 10 confirmed recurrent novae in the Galaxy with recurrency times between 10 and 80~years approximately \citep{Schaefer2010ApJS..187..275S}. This number could, however, be larger since other systems with very massive WDs could also be recurrent, they have simply not yet been identified as so. The next imminent eruption is that of T CrB, a symbiotic nova which shows a recurrence of about 80 years \cite{2023MNRAS.524.3146Schaefer_TCrB} and for which the next explosion is calculated to happen on\mbox{ 2024.4 ± 0.3 \citep{2023ATel16107....1S_TCrB}}. The latest reports indicate that the source entered the so-called pre-eruption dip and its B and V magnitudes are slowly decreasing, as seen in Figure~\ref{fig_TCrB}. The accretion disk reached a minimum in August--September 2023 and it is showing a fast rebrightening \citep{Munari2024ATel16404....1M}. T CrB is closer than RS Oph (0.9 pc vs. 2.4 pc) and it is expected to reach a flux of about 10 times larger than RS Oph. Its peak optical magnitude is can reach magnitude 2.9 as in previous eruptions, being one of the brightest novae observed. 

HE emission from classical novae was an unexpected discovery, due to the low density of the surrounding environment, but they are now the most frequently detected type of novae. Whether classical novae can emit at VHE is still an open question, although if shocks operate similarly in symbiotic and classical systems, then we could potentially expect VHE emission also from these systems.

\subsection{Microquasars}
\begin{figure}[H]
\centering
\includegraphics[width=10cm]{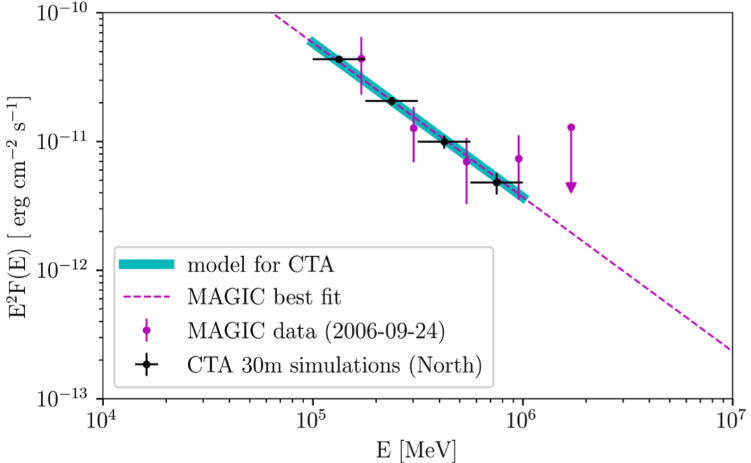}
\caption{Simulated SED of Cyg X-1 as seen by CTA-North (black points) during a flaring episode similar to that reported in \cite{Albert2007} (magenta points). Reprinted from \citep{astronet2021arXiv210603621B} .\label{fig_cygx1}}
\end{figure}

The improved sensitivity of CTA will likely lead to the detection of TeV transient emission from flaring microquasars. When extrapolating the Cygnus X-1 hint observed by~\citep{Albert2007} in the VHE regime, we see that the CTA northern array will be able to detect a similar flare with high significance in only 30 min of observation (see Figure~\ref{fig_cygx1}) \citep{astronet2021arXiv210603621B}. We can expect the future CTA observatory to detect transient emission from other microquasars, probing particle acceleration in jets.

\subsection{Supernovae}
As discussed in Section \ref{sec_sne}, the CTA observatory is expected to detect CCSNe up to a few~Mpc of distance. According to \citep{Sarmah2023arXiv230708744S}, an SN like the recent type II-P SN 2023ixf should be detectable by future experiments such as the CTA even at 7--10~Mpc. In the case Galactic CCSNe, there is the open possibility that current IACTs could detect a VHE counterpart. Since neutrino bursts take place during a core-collapse event, they are expected to precede the electromagnetic radiation from the SN when reaching the Earth, as it happened in SN 1987A. Hence, neutrino bursts are good alert trigger systems for a Galactic CCSNe event, which are rare events in our Galaxy. However, it is expected that the VHE emission is absorbed due to $\gamma$$\gamma$ annihilation during the first 7--10 days, approximately. It can be worth trying to catch the VHE counterpart during the first hours after explosion, since models do not manage to simulate the expected gamma-ray emission so early on and since an observation like this will definitely help constrain the theoretical scenarios for such unique~events.

\subsection{Crab Nebula Flares}
\begin{figure}[H]
\centering
\includegraphics[width=14cm]{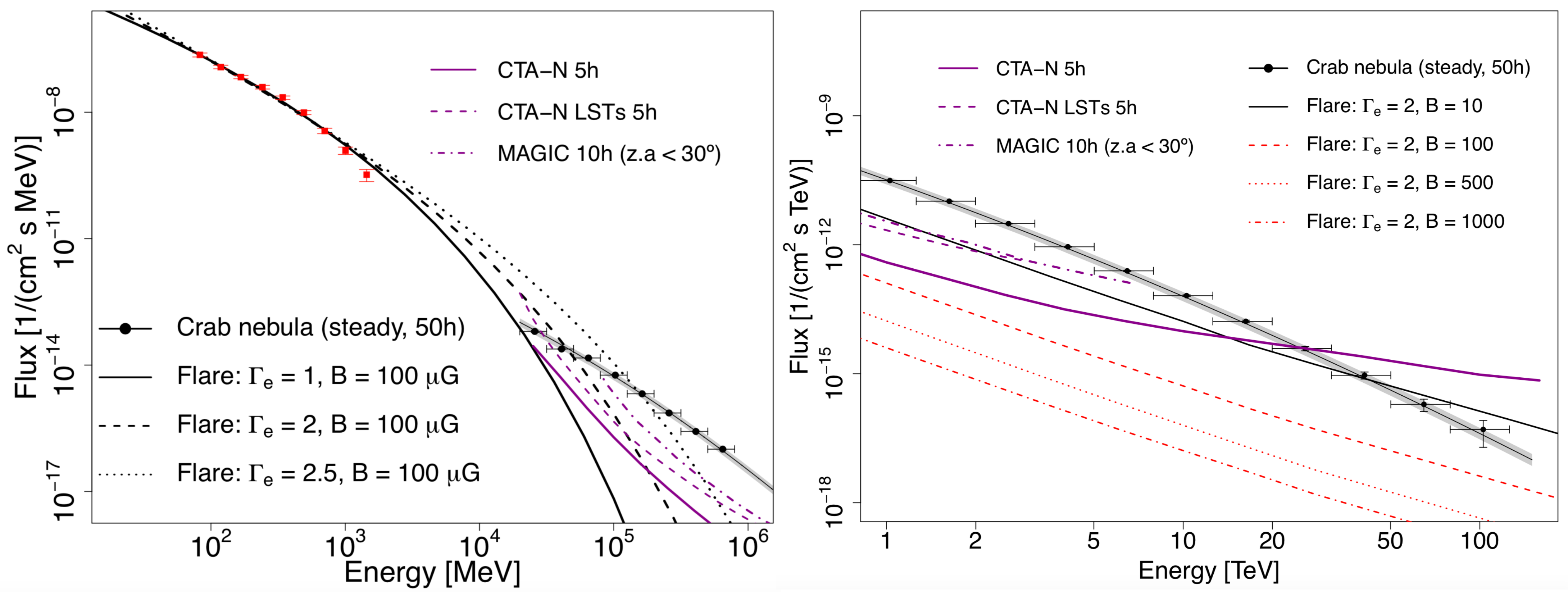}
\caption{Simulated SEDs of the Crab Nebula during different flares compared to steady state. (\textbf{Left})~Synchrotron regime (\textbf{Right}) IC component. The 5 h sensitivity CTA-North array (bold line), the 5 h sensitivity of the 4 LSTs of CTA-North (dashed line), and the 10 h sensitivity of the MAGIC telescopes (dotted--dashed line) are represented. Figure rearranged and reprinted with permission from \citep{2021Mestre}.}  
\label{fig_crab}
\end{figure}

Next-generation instrumentation such as the CTA observatory will count with an increased sensitivity\endnote{\url{https://www.cta-observatory.org/science/ctao-performance/} (accessed on 27 March 2024).} to short timescale transient events~\citep{ALO2023hsa..conf..159L}. It has been explored how the northern array of the CTA observatory will be sensitive to flaring emission from the Crab Nebula \citep{2021Mestre, ALO2022icrc.confE.784L}. The high sensitivity of the array will likely allow for the detection of both the synchrotron end at low energies in few ($\le$5) hours in the case of hard synchrotron flares for magnetic fields with similar or larger intensity than that of the nebula. Even current facilities such as MAGIC could potentially detect bright flares (similar to that of 2011) or at least set strong constraints (see Figure~\ref{fig_crab}).  In the case of the IC component, TeV emission could be detectable if the energy of the electrons is boosted and under certain scenarios, such as soft spectra and mG magnetic fields (right panel of Figure~\ref{fig_crab}).

\subsection{FRB and Magnetars}

The nature of FRBs represents one of the most enigmatic (and recent) hot topics in time-domain astrophysics. The discovery of the association within an FRB-like emission and the Galactic magnetar SGR~1935$+$2154 provided possible evidence about the origin of these events. Magnetars already triggered the interest of IACT at VHE in the search for persistent emission;~see, e.g.,~\citep{2013A&A...549A..23A} and more recently as transient sources as they may undergo important GFs: rare and brief ($\sim$$0.1$~s) bursts of hard X-rays and soft gamma rays, recently detected up to the~GeV range~\citep{2021NatAs...5..385F}. The energy release of a GF may be remarkable, reaching a total value of $10^{44}\div10^{46}$~erg. Although many theoretical models do not envisage magnetars as VHE emitters during their quiescent state, the possibility of having VHE emission during flaring episodes cannot be ruled out. The April 2020 flaring activity of SGR~1935$+$2153 gathered an exceptional extended multi-wavelength coverage, mainly thanks to the above-mentioned FRB connection. The observed X-ray activity showed a harder spectra with respect to the typical bursts from SGR~1935$+$2154 (and other magnetars) although its intensity was relatively moderate and significantly too faint to be classified as a GF. Observations by H.E.S.S. and MAGIC ruled out possible extended emission up to the VHE band for this event~\citep{2021ApJ...919..106A,Lopez-Oramas:2021zd}. Very recently, a candidate magnetar GF from the nearby galaxy M82 has been followed up by MAGIC with a (preliminary) non-detection at VHE~\citep{2023GCN.35068....1P}. 

The high sensitivity to short timescale signals foreseen for the CTA observatory will make it a perfect instrument to magnetars flaring activity follow-up. Furthermore, the new radio facilities that will operate at the time of the CTA will provide the detection of up to hundreds of FRBs per day. Many of these will have good localizations and will be inside the CTA field of view, making it possible to search for prompt and/or delayed VHE emission corresponding with radio activity, unveiling the still-puzzling connection between FRB and~magnetars. 

\subsection{GRBs and GWs} 
\begin{figure}[H]
\includegraphics[width=13cm]{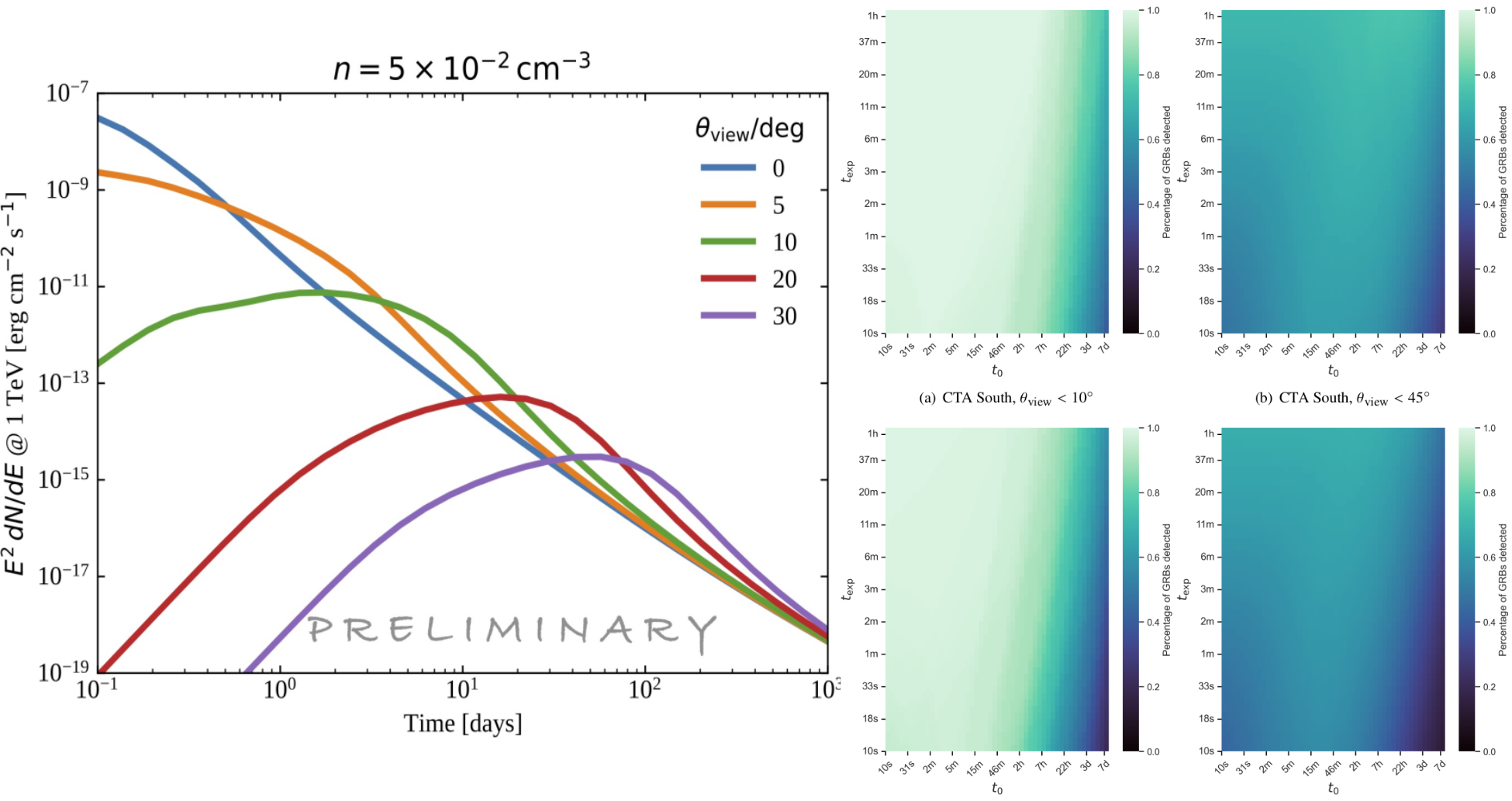}
\caption{({\bf Left}) 1 TeV light curve expected for a GW~170817-like event under more favorable conditions. The circumburst medium density is fixed to ($5 \times 10^{-2}$~cm$^{-3}$), while different viewing angles are plotted. Reprinted with permission from~\citep{2022icrc.confE.944S}. ({\bf Right}) Detectability of sGRBs with CTA-South (upper panels) and CTA-North (lower panels) array for the simulated events given latency and exposure time. The left panels show a subset of the sources with viewing angle $<$$10^{\circ}$, while the right panels show all sources with view $<$$45^{\circ}$. Reprinted with permission from~\citep{2023arXiv231007413G}.} 
\label{fig:gwprospects}
\end{figure}

Whether the sources of GWs are BNS merger or CCSNe, electromagnetic emission up to the VHE may be envisaged. Expectations for VHE emission from CCSNe likely pose these sources out of reach for current IACTs (see~Section~\ref{sec_sne}). On the other hand, although challenging, the VHE counterpart of BNS mergers stands in a better chance of detection for running facilities. The link between sGRB and GWs has indeed been proven by the detection of GW/GRB~170817A, while a hint for VHE also from sGRB (long GRBs are now known to be VHE emitters) has been achieved by MAGIC in the case of the short GRB~160821B~\citep{2021ApJ...908...90A}. Thus, it is justified to assume that each BNS merger may result in an sGRB launching a relativistic jet. However, GW~170817, the only event with a firmly detected electromagnetic counterpart, did not show any hint of~GeV-TeV emission and detailed emission models for this event do not foresee a VHE component strong enough to be detectable with current IACTs~\citep{2022icrc.confE.944S}. Regardless, such negative prospects have to be considered as not conclusive. Within the framework of an off-axis GRB as the source of electromagnetic radiation in a BNS merger, geometry plays a key role in the expected emission at all wavebands. In the case of GW~170817, the relatively large viewing angle of $\theta$$\sim$$15^{\circ}\div25^{\circ}$ played a key role in suppressing the VHE emission component. Viewing angles closer to an on-axis geometry may certainly increase the flux expected at VHE although anticipating the peak time of the emission. Furthermore, the circumburst conditions may also have a significant impact on the expected spectrum. The low interstellar medium density for GW~170817 ($10^{-4}$~cm$^{-3}$) stands as a disadvantage for a detectable VHE signal. In~\citep{2022icrc.confE.944S}, an example of light curves at 1 TeV for a jet with the same parameters as that of GW~170817, but with a denser circumburst medium ($5 \times 10^{-2}$~cm$^{-3}$) (Figure~\ref{fig:gwprospects} left plot).   

It is important to remark that one of the keys for a successful GW follow-up lies in the synergies with other facilities and in the optimization of the observing strategy. As a matter of fact, GRB~180817A was a sub-threshold event, several orders of magnitudes less luminous than a standard GRB, although located much closer to us compared to the average GRB population. Hence, this event may not have been followed up if no gravitational wave was detected. The extensive multi-waveband follow-up has proved to be the key to identify the counterpart and its nature, representing a takeaway message for future observations. Within this framework, a large effort is currently taking place within the CTA consortium to optimize VHE follow-up strategy for near future observations (LVK run O5, planned for 2027~\endnote{\url{https://observing.docs.ligo.org/plan/} (accessed on 27 March 2024).}), where the parallel operation of more GW interferometers will allow for the localization of new merger events with much better precision.

In~\citep{2023arXiv231007413G}, a preliminary estimation of CTA detection capabilities on GWs/GRBs is reported. A set of simulated BNS mergers and their associated GW signals~\citep{2022ApJ...924...54P} are used taking into account realistic astrophysical distributions of masses, spins, distances, and sky locations of the neutron stars. Each merger is associated with a simulated sGRB. The authors use an empirical approach that does not need to assume any specific particle population or radiative process for the production of gamma-rays according to the empirical evidence collected by IACT observations of GRBs. Then, the luminosity in the TeV range is assumed to be comparable to the one at lower energies (in the soft X-ray range), and the spectra are assumed to have a photon index around $-$2.2. The synthetic spectra are then analyzed by means of CTA analysis tools and Instrument Response Functions (IRFs). The estimation of the integration time required to achieve a detection with CTA is reported in Figure~\ref{fig:gwprospects} (right panel) as a function of the time needed by the telescopes to point at the region of interest.

It is clear that the CTA will represent a unique instrument to achieve a VHE detection of a GW counterpart, shedding light into the physics of GRB and BNS mergers dynamics and setting a key step for the future multi-messenger astronomy.

\section{Conclusions}
\label{sec_conc}

The past decade marked the beginning of the era of multi-messenger observations accompanied, in parallel, by the remarkable development of time-domain astronomy. In the~GeV-TeV energy range, in particular, new sources of VHE gamma rays have been identified, opening new perspectives for transient astrophysics in this energy regime.

One of the newly identified class of Galactic VHE emitters are novae thanks to the recent discovery of VHE signal of hadronic origin in the recurrent symbiotic nova RS Oph. These novae create bubbles of enhanced cosmic-ray density in their close environment at $\sim$pc scales. Other (recurrent) symbiotic systems such as T CrB are expected to be detected by the current generation of IACTs in the very near future. The discussion on whether classical novae are TeV emitters is still open and will hopefully be addressed over the next years.

At extragalactic distances, the detection of the TeV counterpart of GRBs was finally achieved by current IACTs after a quest which lasted for more than 20 years. TeV GRBs were first detected in 2018--2019 and since then a total of four long GRBs located at redshifts between 0.0785 and 1.1 have been reported by IACTs during the afterglow phase. Furthermore, the brightest GRB of all times, GRB~221009A, has been recently detected by LHAASO above the 10 TeV, opening new possibilities for GRB study with instruments not originally thought for GRB follow-up due to their relatively large energy threshold such as particle array detectors or the ASTRI Mini-Array. The hint of detection at VHE from the short-GRB~160821 also proved the possible link between VHE and the GW emission from BNS mergers. In the near future, thanks to the improved sensitivity in the GW interferometers and the new-generation IACTs,  follow-up observations of GW + VHE will connect the gamma-ray emission with the formation and evolution of the GW-central engine, shedding light into the physics of these extreme cosmic events. 

New intriguing transients such as FRBs and their connection with magnetars represent a very recent development and a still marginally explored field for VHE transient astrophysics. Although magnetars \textit{per se} are not found to be steady gamma-ray emitters, they have been detected in the~GeV range by {\it Fermi}-LAT during giant flare episodes. Furthermore, magnetar-based models predict emission up to the VHE correlated in time with FRBs. The higher sensitivity to short IACTs compared to space-based instruments represents a unique feature for exploring the wide and complex range of transient phenomenology embedded in the magnetar--FRB scenario.

The current generation of IACTs is still on the catch of other transient events that are known HE emitters such as the enhanced flaring emission from PWNe, notably the Crab Nebula or flaring (massive) microquasars as, e.g., Cygnus X-1 or Cygnus X-3. Other transient phenomena are still elusive both in the HE and VHE regimes, such as core-collapse SNe (despite some candidate associations at HE), the VHE counterpart of kilonovae from GWs, or TDEs. The identification of any of these sources as TeV emitters will undoubtedly push the boundaries of our knowledge and open new research areas. In this regard, future instrumentation such as the CTA observatory, with enlarged energy range, improved (short timescale) sensitivity, and fast response capabilities will set new frontiers in time-domain TeV astrophysics. 


\vspace{6pt} 


\authorcontributions{The authors contributed equally to this work. All authors have read and agreed to the published version of the manuscript.}

\funding{This research is part of the project RYC2021-032991-I, funded by MICIN/AEI/10.13039/ 501100011033, and the European Union “NextGenerationEU”/PRTR. }

\dataavailability{Not applicable.}

\conflictsofinterest{The authors declare no conflicts of interest.}



\abbreviations{Abbreviations}{
The following abbreviations are used in this manuscript:\\
\noindent 
\begin{tabular}{@{}ll}
BH & Black hole\\
CCSNe & Core-collapse supernovae\\
FRB & Fast radio burst\\
GRB & Gamma-ray burst\\
GW & Gravitational wave\\
HE & High energy\\
IACT & Imaging Air Cherenkov Telescopes \\
IC & Inverse Compton\\
NS & Neutron star\\
RG & Red giant\\
SSC & Synchrotron self Compton\\
SED & Spectral energy distribution\\
sGRB & Short gamma-ray burst\\
SNe & Supernovae\\
TDE & Tidal disruption event\\
UL & Upper limit\\
VHE & Very-high-energy\\
WD & White dwarf\\
\end{tabular}
}

\begin{adjustwidth}{-\extralength}{0cm}
\printendnotes[custom] 

\reftitle{References}


\PublishersNote{}
\end{adjustwidth}
\end{document}